\documentclass[12pt]{article}
\usepackage{eurosym}
\usepackage{graphicx}
\usepackage[utf8]{inputenc}
\usepackage[T1]{fontenc}
\usepackage{indentfirst}
\usepackage[margin=0.6in,nomarginpar]{geometry}
\usepackage[final]{hyperref}
\usepackage{amsmath}
\usepackage{hyperref}
\usepackage{cite}
\usepackage{subcaption}
\usepackage{caption}
\usepackage{amssymb}
\usepackage{multirow}
\usepackage[table]{xcolor}
\usepackage{orcidlink}
\usepackage{color}
\usepackage{float}
\usepackage{multirow}
\usepackage{colortbl}
\hypersetup{
colorlinks=true,
linkcolor=blue,
citecolor=blue,
filecolor=magenta,
urlcolor=blue
}

\begin{document}

\title{Schwarzschild black hole surrounded by a cloud of strings in the background of perfect fluid dark matter}

\author{B. Hamil 
 \orcidlink{0000-0002-7043-6104} \thanks{%
hamilbilel@gmail.com} , \\
Laboratoire de Physique Math\'{e}matique et Subatomique,\\
Facult\'{e} des Sciences Exactes, Universit\'{e} Constantine 1, Constantine,
Algeria. \and B. C. L\"{u}tf\"{u}o\u{g}lu 
 \orcidlink{Orcid ID : 0000-0001-6467-5005} \thanks{%
bekir.lutfuoglu@uhk.cz 
 (Corresponding author)} , \\
Department of Physics, Faculty of Science, University of Hradec Kralove, \\
Rokitanskeho 62/26, Hradec Kralove, 500 03, Czech Republic. }
\date{\today }
\maketitle

\begin{abstract}
This manuscript investigates a Schwarzschild black hole surrounded by perfect fluid dark matter embedded in a cloud of strings. The effects of its surroundings on thermodynamics, timelike and null geodesics, shadows, and quasinormal modes are analyzed. It is demonstrated that changes in spacetime, induced by the surrounding environment, significantly influence the stability, thermal phases, energy dynamics, particle trajectories, and observable features of the black hole's shadow, as well as the oscillation frequency and decay rate.
\end{abstract}


%EndAName

\section{Introduction}

The prediction of black holes, one of the most profound and enigmatic outcomes of Albert Einstein’s General Theory of Relativity, has fascinated scientists for decades. This theoretical concept became an observational reality with the groundbreaking work of the Laser Interferometer Gravitational Wave Observatory (LIGO) collaboration, which not only detected gravitational waves that confirm a key aspect of General Relativity \cite{Abbott1}, but also recorded the unprecedented merger of two black holes \cite{Abbott2}. Further validating these predictions, the Event Horizon Telescope (EHT) collaboration captured the first-ever image of a black hole’s shadow in the galaxy Messier 87 (M87*) \cite{Akiyama2019}, and later, imaged Sagittarius A*, the supermassive black hole at the center of the Milky Way \cite{Akiyama2022}. These extraordinary discoveries by LIGO and EHT have revolutionized astrophysics, not only opening new frontiers in black hole research but also providing a foundation for detailed analyses of their fundamental properties.

One of the key features in black hole imagery is the shadow it casts \cite{Chen2023}. This shadow appears as a two-dimensional dark area in the sky, formed by light rays bent around the black hole’s event horizon due to the extreme curvature of spacetime \cite{Synge1966}. The size and shape of the shadow provide crucial information about the black hole's mass, spin, and the surrounding spacetime geometry \cite{Bardeen1973}. For instance, a Schwarzschild black hole (SBH), which is non-rotating and spherically symmetric, forms a perfectly circular shadow. In contrast, a Kerr black hole, characterized by its rotation, produces an asymmetric and elongated shadow due to the frame-dragging effect of its spin. The study of these shadow shapes has become a critical observational tool for understanding black hole properties, fueling a rapid increase in related research literature; examples of such studies can be found in
\cite{Tsu1, Tsu2, Tsu3, WeiMann, Babar, Kumar, Singh, Nodehi2020, LiGuo, Zhang21, Thomas1, AtamuratovJamil2021, Anacleto2021, Rayimbaev2022, Das, Pantig2022, Campos2022, GuoWD, PG1, ZG1, Sunny1, Sunny2, Sunny3, Sunny4, Sunny5, Sunny6,  Vir1, Biz2023, Biz2023b, Du, Akhil, Kara, Ali1, Chowdhuri, Molla, Olmo, AtamuratovJamil2023, Hoshimov2024, Singh2024, Li2024, Campos2024, Kon2024a}.

The environment around a black hole, such as the presence of dark matter, can significantly affect the characteristics of its shadow. Studies have shown that when a black hole is surrounded by dark matter, the distribution of this matter can influence the shadow’s size and shape. For example, black holes with dark matter halos exhibit shadows that can either increase or decrease in size depending on the density and proximity of the dark matter to the black hole. If the dark matter is highly concentrated near the black hole, it could distort the spacetime and thus alter the shadow’s appearance, making it more detectable. In particular, the photon orbits (or the paths light takes around the black hole) can be altered by the presence of dark matter, leading to variations in the shadow’s intensity and geometry compared to a vacuum SBH \cite{Konoplya2019, Pantig2020, Saurabh2021, Sergio2023, Malligawad 2024, Wu2024, Rizwan2024}.

In string theory, the concept that fundamental particles are replaced by one-dimensional strings has led to new perspectives on black holes, particularly with the idea that a cloud of strings (CoS) could serve as a source of the gravitational field. Letelier first proposed this model, providing an exact solution for an SBH surrounded by a CoS \cite{Letelier1979}. Since then, further research has explored the impact of these string clouds on various black hole solutions \cite{Toledo2019, Costa2019, Chabab2020, Cardenas2021, Mustafa2021, He2022, Mustafa2022, Atamurotov20221, Manuel2022,  Liu2023, Zahid2023,  Yang2024, Rani2024, Sudhanshu2024, Chaudhary2024, Badawi2024}. %For example, Toledo et al. in \cite{Toledo2019} studied the thermodynamics of a Reissner–Nordström black hole surrounded by a quintessence matter field and a cloud of strings, while Costa et al. in \cite{Costa2019} focused on the same aspects for a Schwarzschild black hole. Subsequently, Cárdenas et al. explored the astrophysical implications of a Schwarzschild black hole associated with a cloud of strings and quintessence through standard general relativistic tests conducted in the solar system, revealing potential modifications to gravitational interactions and celestial dynamics \cite{Cardenas2021}.  Additionally, the authors explored the motion of photons and test particles \cite{Mustafa2021}, as well as the shadow, photon sphere \cite{He2022}, and gravitational weak lensing \cite{Mustafa2022} in a Schwarzschild black hole influenced by quintessence and string clouds, highlighting their impact on black hole observations and behavior. In \cite{Manuel2022}, the authors showed that the Bardeen black hole becomes singular in the presence of a string cloud, while the Simpson–Visser solution remains regular. Soon after, it was found that the quasinormal modes of the Bardeen black hole are significantly affected by the presence of a string cloud \cite{Liu2023}, and its thermodynamic properties and geometry are also altered \cite{Rani2024}. Moreover, studies by Yang et al. demonstrated that the string cloud parameter plays a crucial role in shaping the shadow of rotating black holes, influencing black hole mimicry and the behavior of quasinormal modes \cite{Yang2024}. Recently, in \cite{Badawi2024} and \cite{Chaudhary2024}, the authors explored the effects of GUP and EGUP corrections on the images and stability of black holes with a cloud of strings, respectively.
For example, Toledo et al. \cite{Toledo2019} studied the thermodynamics of a Reissner–Nordström black hole surrounded by a quintessence matter field and a CoS, while Costa et al. \cite{Costa2019} focused on similar aspects for a SBH. Subsequently, Cárdenas et al. \cite{Cardenas2021} explored the astrophysical implications of a SBH associated with a CoS and quintessence through standard general relativistic tests conducted in the solar system, revealing potential modifications to gravitational interactions and celestial dynamics. Additionally, the authors examined the motion of photons and test particles \cite{Mustafa2021}, as well as the shadow, photon sphere \cite{He2022}, and gravitational weak lensing \cite{Mustafa2022} in a SBH influenced by quintessence and string clouds, highlighting their impact on black hole observations and behavior. In \cite{Manuel2022}, Rodrigues et al. demonstrated that the Bardeen black hole becomes singular in the presence of a string cloud, while the Simpson–Visser solution remains regular. Soon after, it was found that the quasinormal modes of the Bardeen black hole are significantly affected by the presence of a string cloud \cite{Liu2023}, along with alterations to its thermodynamic properties and geometry \cite{Rani2024}. Moreover, studies by Yang et al. \cite{Yang2024} illustrated that the string cloud parameter plays a crucial role in shaping the shadow of rotating black holes, influencing black hole mimicry and the behavior of quasinormal modes. Recently, in \cite{Badawi2024} and \cite{Chaudhary2024}, the authors investigated the effects of quantum corrections raised by generalized- and extended generalized- uncertainty principles on the images and stability of black holes with a CoS, respectively.

%The perfect fluid dark matter (PFDM) model offers an intriguing framework for understanding the interplay between dark matter and black holes. Unlike standard particle-based dark matter models, the PFDM approach treats dark matter as a continuous, non-viscous fluid, characterized by specific equations of state that influence its dynamics. This model has been applied to the study of black holes, particularly in scenarios where dark matter may cluster around these massive objects, altering their observable properties. For instance, the PFDM model predicts distinct modifications to black hole metrics, including changes to the Schwarzschild or Kerr solutions, potentially affecting accretion rates, gravitational wave signals, and event horizon structures. Such deviations could provide new insights into the nature of dark matter and its distribution in galactic halos, offering a novel pathway to reconcile cosmological observations with the physics of black holes. As a result, the PFDM model represents a promising alternative to traditional dark matter theories, particularly in explaining phenomena observed at galactic and cosmological scales.

On the other hand, the need to study dark matter remains critical, as it makes up approximately $27$\%$ $ of the universe's total energy density, yet its true nature continues to be one of the greatest unsolved mysteries in modern physics. Understanding dark matter is essential not only for explaining galaxy formation and large-scale cosmic structures but also for investigating its interaction with dark energy, which drives the accelerated expansion of the universe. In this context, the perfect fluid dark matter (PFDM) model offers an intriguing framework for exploring the role of dark matter, particularly in relation to black holes. Unlike standard particle-based models, the PFDM approach treats dark matter as a continuous, non-viscous fluid governed by specific equations of state that shape its dynamics \cite{Kiselev2002}. This model has been applied to scenarios where dark matter clusters around black holes \cite{Qiao2023}, altering their observable properties. For instance, the PFDM model predicts modifications to black hole metrics, such as changes to the Kerr \cite{Xu2018, Hou2018, Rizwan2019}, Schwarzschild \cite{Rayimbaev2021, BBB1}, Reissner-Nordström \cite{Xu2019}, Bardeen \cite{Zhang2021}, or Euler-Heisenberg \cite{Ma2024, BBB2} solutions, potentially influencing geodesics \cite{Das2021, Das2022, Das2024}, gravitational wave signals \cite{Shaymatov2021, Li2022},  stability and phase transitions \cite{Hendi2020, Kumar2023, Sood2024}, lensing \cite{Atamurotov2021},  accretion discs \cite{Heydarifard2023}, shadows \cite{Haroon2019, Ma2021}, deflection angle \cite{Atamurotov20222}, quasinormal modes \cite{Sapher24}, greybody factors \cite{Sharif2022}, thermodynamics \cite{Abbusattar2023, Ndongmo2023, Rakhimova2023}, and event horizon structures \cite{Feng2024}. These deviations provide new insights into the nature and distribution of dark matter in galactic halos, offering a novel way to reconcile cosmological observations with black hole physics. As a result, the PFDM model presents a promising alternative to traditional dark matter theories, particularly in explaining phenomena at galactic and cosmological scales.

In a recent article \cite{Sood2024}, Sood et al. explored Letelier black holes in AdS spacetime surrounded by PFDM, revealing significant insights into phase transition phenomena, photon orbits, and critical behaviors. Motivated by these findings, we aim to extend the investigation to an SBH immersed in a CoS and PFDM, focusing specifically on the case without the influence of AdS spacetime. Our study will examine the black hole’s thermodynamics, shadows, geodesics, and quasinormal modes, with the goal of understanding how the presence of PFDM and string clouds affects these fundamental properties. Through this exploration, we seek to uncover new insights into the interaction between dark matter, string structures, and black hole dynamics, offering a deeper understanding of their microphysical nature and potential astrophysical signatures.

The manuscript is organized as follows: Section \ref{sec2}  provides an overview of the mathematical framework for the SBH in the presence of a CoS and PFDM. Section \ref{sec3} discusses the black hole's thermodynamic properties, highlighting the effects of the string and PFDM parameters. Section \ref{sec4} explores the geodesic structure, followed by Section \ref{sec5}, which examines the black hole’s shadow in the context of a CoS and PFDM. Section \ref{sec6} investigates the quasinormal modes of the system. Finally, the manuscript concludes with a summary of the key findings. {  It is important to note that throughout the manuscript, natural units are adopted, where $\hbar=k_{B}=G=c=1$.}

\section{SBH immersed in a CoS and PFDM} \label{sec2}

In this section, we aim to derive the Schwarzschild solution for a spacetime surrounded by a CoS in the presence of PFDM. To achieve this, we consider the following action
\begin{equation}
S=\int d^{4}x\sqrt{-g}\left[ \frac{1}{16\pi }R-\mathcal{L}^{PFDM}\right]+S^{CS},  \label{Action}
\end{equation}%
where $g$ is the determinant of the metric tensor $g^{\mu \nu}$ and $R$ denotes the scalar curvature. Here, $\mathcal{L}^{PFDM}$ represents the Lagrangian density for PFDM, and $S^{CS}$ corresponds to the Nambu-Goto (NG) action, which describes string-like objects, as defined in \cite{Letelier1979, Manuel2022}:
\begin{equation}
S^{CS}=\frac{1}{8\pi }\int_{\Pi }\sqrt{-\gamma }\mathcal{M}d\lambda
^{0}d\lambda ^{1}.
\end{equation}
In this action, $\mathcal{M}$ denotes a dimensionless constant that characterizes the string. The parameters $\lambda ^{0}$ and $\lambda ^{1}$ are a timelike and a spacelike parameter, respectively, while $\gamma $ represents the determinant of $\gamma _{ab}$, which is the induced metric on a submanifold defined by  
\begin{equation}
\gamma _{ab}=g_{\mu \nu }\frac{\partial x^{\mu }}{\partial \lambda ^{a}}%
\frac{\partial x^{\nu }}{\partial \lambda ^{b}}.
\end{equation}%
Using a spacetime bivector $\Pi ^{\mu \nu }=\epsilon ^{ab}\frac{\partial
x^{\mu }}{\partial \lambda ^{a}}\frac{\partial x^{\nu }}{\partial \lambda
^{b}}$, we can express the NG action with:
\begin{equation}
S^{CS}=\int_{\Pi }\sqrt{-\frac{1}{2}\Pi _{\mu \nu }\Pi ^{\mu \nu }}\mathcal{M%
}d\lambda ^{0}d\lambda ^{1},
\end{equation}%
where $\epsilon ^{ab}$ is the Levi-Civita symbol, $\epsilon^{01}=-\epsilon ^{10}=1.$ We then vary the total action given in (\ref{Action}) with respect to the metric and obtain
the field equations
\begin{equation}
G_{\mu \nu }=R_{\mu \nu }-\frac{1}{2}g_{\mu \nu }R=\left( T_{\mu \nu
}^{CS}+8\pi T_{\mu \nu }^{PFDM}\right) .  \label{3}
\end{equation}%
In this formulation, the energy-momentum tensor for the CoS is expressed as 
\begin{equation}
T^{\mu \nu CS}=\mathcal{M}\frac{\Pi ^{\mu \sigma }\Pi _{\sigma }^{\text{ }
\nu }}{\sqrt{-\gamma }}, \label{rev1}
\end{equation}
where the only non-vanishing component of the bivector $\Pi ^{01}=-\Pi ^{10}$ is a function of the radial coordinate. This component is explicitly written as
\begin{equation}
\Pi ^{01}=\sqrt{-\gamma }=\frac{a}{\mathcal{M}r^{2}},  \label{2}
\end{equation}
leading to 
\begin{equation}
    {T_{0}^0}^{CS}=-\frac{a}{r^{2}},
\end{equation}
where $a$ is an integration constant associated with the strings, constrained within the range $0<a<1$. 

On the other hand, the second tensor, which represents the energy-momentum tensor for the PFDM, is written in the conventional orthonormal reference frame as follows \cite{Yang2012}:
\begin{equation}
T_{\mu }^{\nu PFDM}=diag\left( -\rho ,p_{r},p_{\theta },p_{\varphi }\right) . \label{rev2}
\end{equation}%
Here, $\rho $ denotes the energy density and is expressed in terms of the PFDM constant $\alpha$ as 
\begin{equation}
\rho =-\frac{\alpha }{8\pi r^{3}}, \label{1}
\end{equation}%
and the remaining components correspond to the radial and angular pressures. 

Now let us consider a spherically symmetric spacetime, characterized by the metric function $f(r)$, which depends on the radial coordinate $r$, and is given by the form
\begin{equation}
f\left( r\right) =1-\frac{2m\left( r\right) }{r},  
\end{equation}
where $m(r)$ is the mass function. Then, we employ the corresponding line element 
\begin{equation}
ds^{2}=-f\left( r\right) dt^{2}+\frac{1}{f\left( r\right) }dr^{2}+r^{2}d\theta ^{2}+r^{2}\sin ^{2}\theta d\varphi ^{2},  \label{fimetric}
\end{equation}
in conjunction with {Eqs. (\ref{rev1}) and (\ref{rev2}).} Substituting these expressions, we find that the time component of Eq. (\ref{3}) simplifies to
\begin{equation}
-\frac{2}{r^{2}}\frac{d}{dr}m\left( r\right) =-\frac{a}{r^{2}}+\frac{\alpha 
}{r^{3}} .  \label{12}
\end{equation}
We then integrate Eq. (\ref{12})  from $r$ to $\infty $, obtaining the metric function for the SBH in the presence of the CoS and PFDM as follows:
\begin{equation}
f\left( r\right) =1-a-\frac{2M}{r}+\frac{\alpha }{r}\ln \frac{r}{\left\vert
\alpha \right\vert }.\label{bh}
\end{equation}%
It is worthwhile to mention that this metric function is the same as the one given by Sood et al. in  \cite{Sood2024} in the absence of the AdS spacetime term.  Moreover, for  $\alpha =0$, the metric function, which we obtained in Eq. \eqref{bh},  reduces to that of the SBH with a CoS background \cite{Letelier1979}. In the case where $a=0$,  the metric function describes the SBH immersed in PFDM \cite{BBB1}. In the scenario where both $\alpha $ and $a$ are zero, the metric function reduces to the traditional metric function of the SBH.

\begin{figure}[htb!]
\begin{minipage}[t]{0.5\textwidth}
        \centering
        \includegraphics[width=\textwidth]{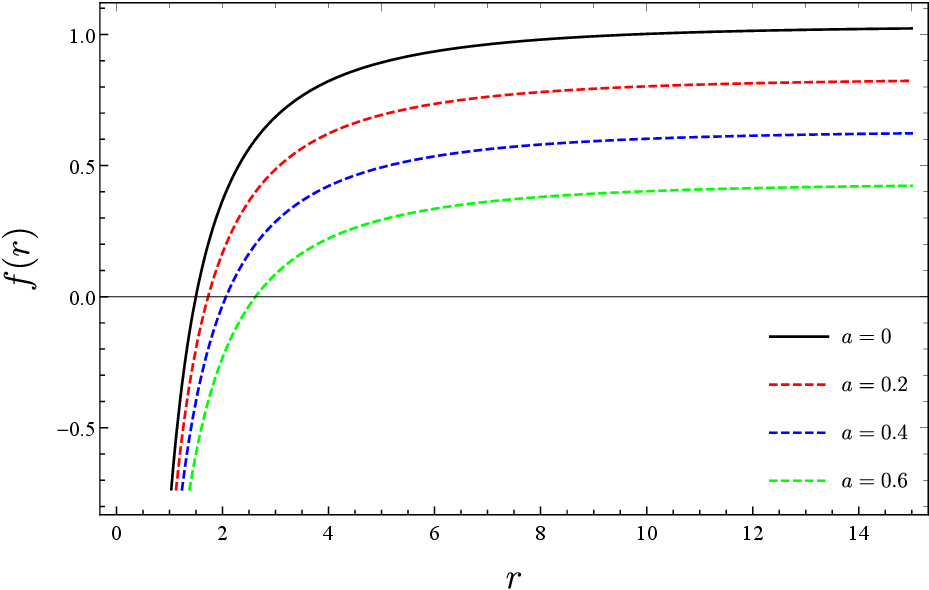}
                \subcaption{$\alpha=0.8$}
        \label{fig:mc}
\end{minipage}
\begin{minipage}[t]{0.5\textwidth}
        \centering
        \includegraphics[width=\textwidth]{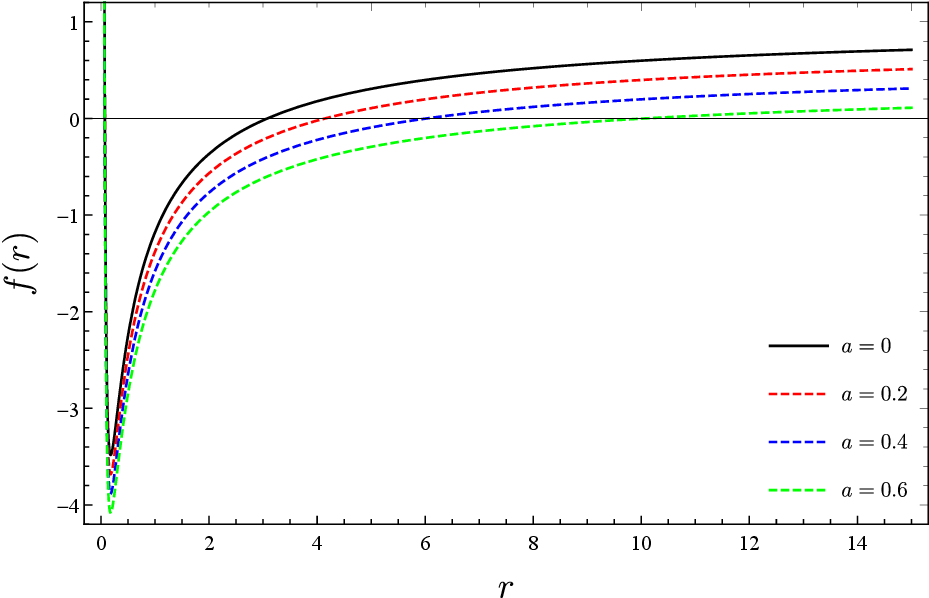}
                \subcaption{$\alpha=-0.8$}
        \label{fig:ma}
\end{minipage}%$
\caption{The effect of the CoS on the $f(r)-r$ diagrams of the SBH surrounded by PFDM.}
\label{fig:ff}
\end{figure}

To examine the impact of the CoS and PFDM on the black hole, we plot the metric function versus the radius diagrams in Fig.\ref{fig:ff}. { Here, we consider both positive and negative values of $\alpha$ in the remainder of our analysis to explore the full range of possible effects on black hole properties.}
\newpage

\noindent {
In both scenarios, increasing the CoS parameter causes a downward shift in the metric function, thereby influencing the horizon's location. For $\alpha = 0.8$, as the CoS parameter increases, the metric function reaches positive values at progressively larger radial distances, effectively moving the horizon outward. In contrast, a distinct structural configuration emerges in the scenario with $\alpha = -0.8$. As shown in Fig.~\ref{fig:ma}, the metric function displays two distinct horizons: event horizon $r_{H}$ and an inner Cauchy horizon $r_{C}$, in contrast to the single horizon observed in Fig.~\ref{fig:mc}. This result indicates that the PFDM parameter has a substantial impact on spacetime curvature, and consequently, on the black hole’s thermal properties, particularly in relation to its stability. Finally, as $r$ increases, the metric function approaches a constant value in both cases, with the influence of $a$ and $\alpha$ diminishing at larger radii.
}

{
Upon comparison with the existing literature, we observe similar results. For instance, Al-Badawi et al. \cite{Badawi2024}, who investigated the GUP-corrected black holes in the presence of a CoS, found that increasing the CoS parameter values leads to a larger horizon radius, a trend that also holds in our study for both positive and negative PFDM scenarios.}

\section{Thermodynamics} \label{sec3}

In this section, we discuss the thermodynamic characteristics of the black hole modeled by Eq. (\ref{bh}). We start by calculating the mass of the black hole using the condition $f(r_H) = 0$, where $r_H$ is the event horizon radius. This leads to the following expression for the black hole mass:
\begin{equation}
M=\frac{r_{H}}{2}\left( 1-a+\frac{\alpha }{r_{H}}\ln \frac{r_{H}}{\left\vert
\alpha \right\vert }\right). \label{mass}
\end{equation}
For $a=0$, Eq. (\ref{mass}) simplifies to 
\begin{equation}
M=\frac{r_{H}}{2}\left( 1+\frac{\alpha }{r_{H}}\ln \frac{r_{H}}{\left\vert
\alpha \right\vert }\right), \label{massf}  
\end{equation}
which describes the mass of the SBH surrounded by PFDM \cite{BBB1}. Instead, if one sets $\alpha=0$, then Eq. (\ref{mass}) reduces to the mass of the SBH in the presence of CoS \cite{Letelier1979}:
\begin{equation}
M=\frac{r_{H}}{2}\left( 1-a\right), \label{masscos}  
\end{equation}
and for $\alpha=a=0$, we retrieve the mass of a standard SBH.

The temperature associated with a black hole, referred to as the Hawking temperature, is directly proportional to its surface gravity $\kappa$, with the relationship expressed by the equation $T = \frac{\kappa}{2\pi}$, where the surface gravity $\kappa$ is defined by the expression 
\begin{equation}
\kappa =-\left. \frac{1}{2\sqrt{-g_{00}g_{11}}}g_{00}^{{\prime}}\right\vert _{r=r_{H}}.\label{sur}
\end{equation}%
Using Eqs. (\ref{bh}) and (\ref{sur}), we
derive a relationship between the black hole's temperature and horizon radius in the following form:
\begin{equation}
T=\frac{1}{4\pi r_{H}}\left( 1-a+\frac{\alpha }{r_{H}}\right).\label{temp}
\end{equation}
We note that the Hawking temperature reduces to $T=\frac{1-a}{4\pi r_{H}}$ when $\alpha =0$ and to $T=\frac{1}{4\pi r_{H}}\left( 1+\frac{\alpha }{r_{H}}\right) $ when $a=0$, as previously given in earlier works \cite{Letelier1979, BBB1}, while 
in the absence of both SoC and PFDM, Eq. \ref{temp} yields the usual form, $T=\frac{1}{4\pi r_{H}}$.

\begin{figure}[H]
\begin{minipage}[t]{0.5\textwidth}
        \centering
        \includegraphics[width=\textwidth]{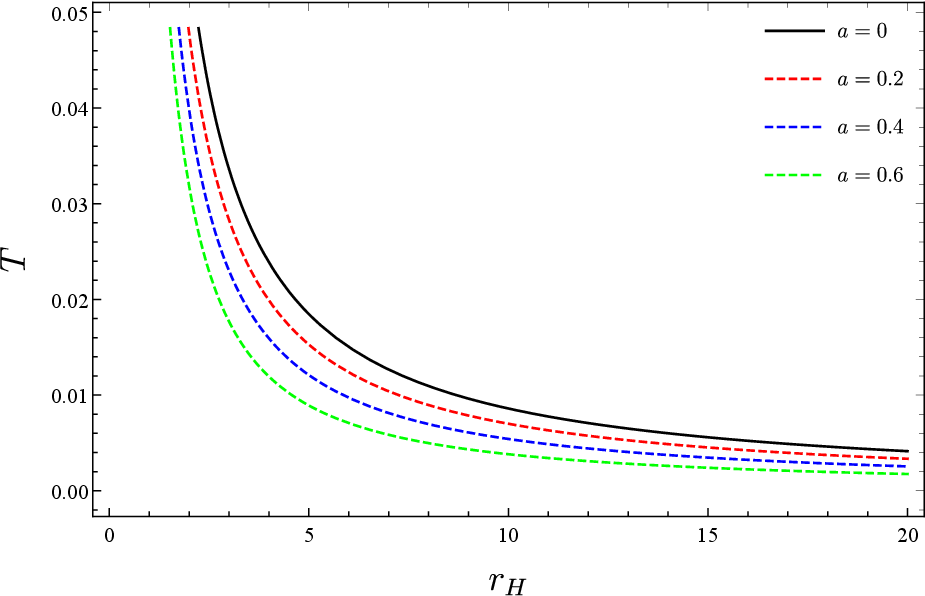}
                \subcaption{$\alpha=0.8$}
        \label{fig:ta}
\end{minipage}%$
   \begin{minipage}[t]{0.5\textwidth}
        \centering
        \includegraphics[width=\textwidth]{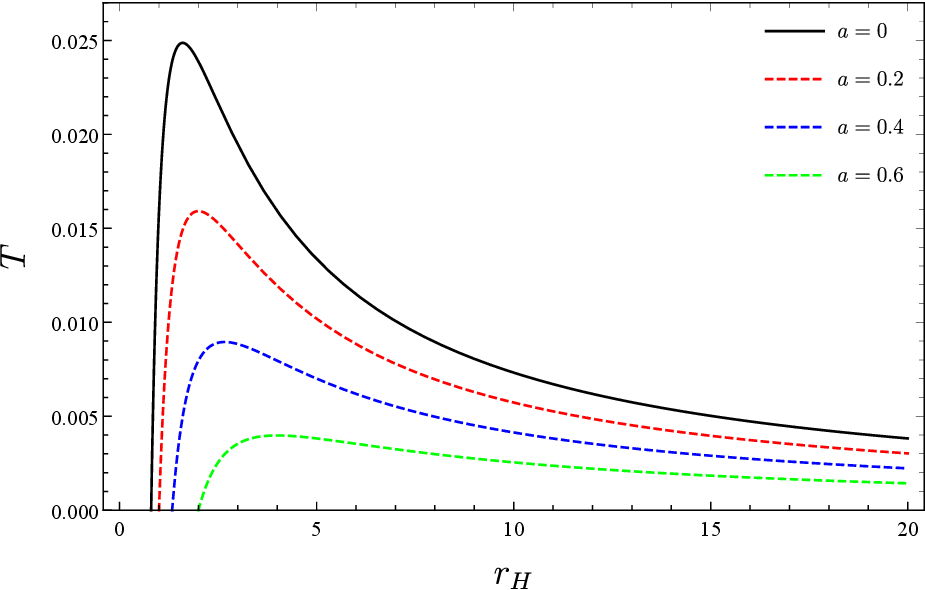}
                \subcaption{$\alpha=-0.8$}
        \label{fig:tc}
\end{minipage}
\caption{Temperature versus horizon radius for different values of the $a$ in two PFDM scenarios.}
\label{fig:tt}
\end{figure}

\noindent In Fig. \ref{fig:tt}, we plot the Hawking temperature of the SBH immersed in PFDM and surrounded by a CoS for two distinct scenarios and various parameter values. In a scenario with a positive PFDM parameter, the temperature indicates that the SBH is cooling down monotonically by the Hawking radiation as its size increases. A comparison between Fig. \ref{fig:ta} and Fig. \ref{fig:tc} reveals that for negative PFDM parameters, the divergence observed with positive PFDM parameters disappears. Moreover, during the evaporation process, the Hawking temperature initially increases, reaching a maximum value of 
\begin{equation}
 T^{\max }=-\frac{\left( 1-a\right) ^{2}}{16\pi \alpha }   
\end{equation}
at the critical horizon radius 
\begin{equation}
r_{crt}=-\frac{2\alpha }{1-a},
\end{equation}
before rapidly dropping to zero at the black hole's minimum size,  
\begin{equation}
r_{min}=-\frac{\alpha }{1-a}.    
\end{equation}
Additionally, as the value of the CoS parameter increases, the overall temperature decreases for a given horizon radius. This effect is consistent across both figures, suggesting that a stronger CoS reduces the Hawking temperature. The temperature tends to converge for large horizon radii, indicating that the effects of the CoS and PFDM become less significant for relatively larger black holes.

We then derive the SBH entropy function by utilizing the formula:
\begin{equation}
dS=\frac{dM}{T}.\label{ent}   
\end{equation}
Using Eqs. (\ref{mass}) and (\ref{temp}), the black hole entropy is expressed in its conventional form \cite{BBB1, Sood2024}
\begin{equation}
S=\pi r_{H}^2 . \label{entf} 
\end{equation}
We note that the CoS and PFDM do not explicitly affect the entropy, which remains linearly proportional to the area. It is worth mentioning that the CoS and PFDM parameters do have a direct impact on the black hole and its entropy through the event horizon.

\begin{figure}[htb!]
\begin{minipage}[t]{0.49\textwidth}
        \centering
        \includegraphics[width=\textwidth]{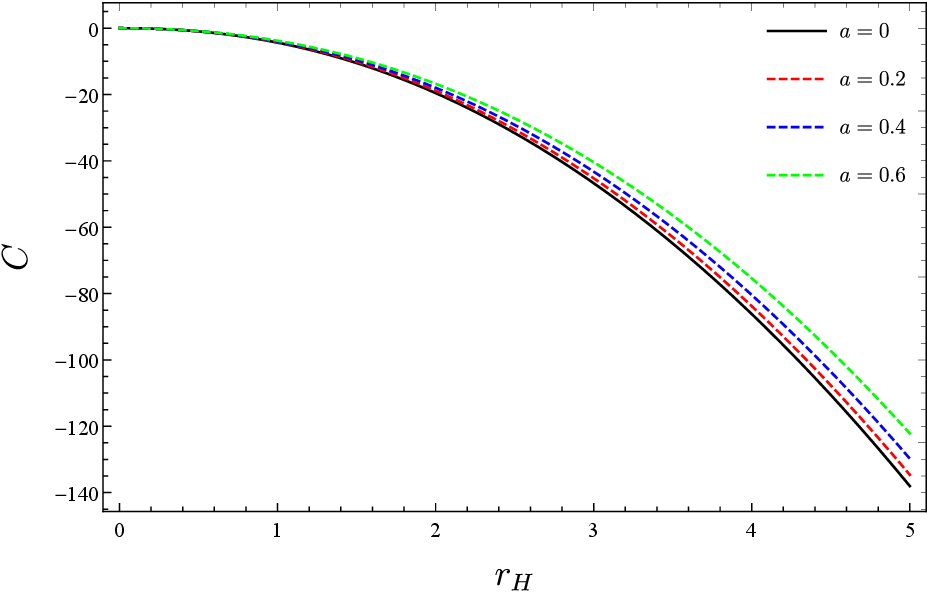}
                \subcaption{$\alpha= 0.8$}
        \label{fig:ha}
\end{minipage}
   \begin{minipage}[t]{0.49\textwidth}
        \centering
        \includegraphics[width=\textwidth]
        {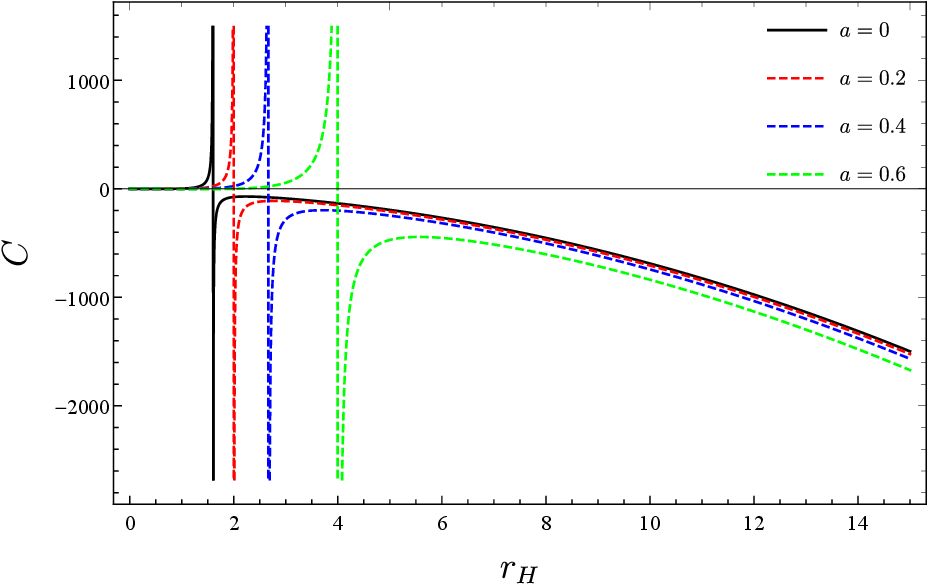}
                \subcaption{$\alpha=-0.8$}
        \label{fig:hc}
\end{minipage}
 \caption{Heat capacity versus horizon radius for different values of $a$ in two PFDM scenarios.}
\label{fig:hh}
\end{figure}

Next, we investigate the thermodynamic stability by utilizing the heat capacity, which is defined as:
\begin{equation}
C=T\frac{\partial S}{\partial T}.\label{heat}
\end{equation}
By substituting Eqs. (\ref{temp}) and (\ref{entf}) into Eq.~\ref{heat}), we obtain
\begin{equation}
C=-2\pi r_{H}^{2}\frac{(1-a)r_{H}+\alpha }{(1-a)r_{H}+2\alpha }.\label{heata}
\end{equation}
It is worth noting that, unlike the entropy function, the heat capacity explicitly depends on both the string cloud and PFDM coefficients. In the absence of these parameters, specifically in the limit $a \rightarrow 0$, Eq. (\ref{heata}) reduces to:
\begin{equation}
C=-2\pi r_{H}^{2}\frac{r_{H}+\alpha }{r_{H}+2\alpha },
\end{equation} 
as given in  \cite{BBB1}. Notably, in the absence of the PFDM parameter $\alpha=0$, Eq. (\ref{heata}) gives the standard form
\begin{equation}
C=-2\pi r_{H}^2,
\end{equation} 
which indicates that the presence of the string cloud does not directly influence the specific heat. To further analyze, we present in Fig. \ref{fig:hh} the plot of the heat capacity for a Schwarzschild black hole surrounded by the CoS and PFDM.

\newpage
\noindent Fig. \ref{fig:ha}  illustrates the behavior of heat capacity with the positive PFDM parameter. The results show that the heat capacity is consistently negative, which signifies the thermodynamic instability of the black hole. Notably, the curves for varying values of $a$ are distinct, particularly for larger values of $r_{H}$. {In} contrast, Fig. \ref{fig:hc} depicts the heat capacity with a negative PFDM parameter, { where} we observe that the heat capacity reaches zero at $r_{H}=r_{min}$, {marking} the minimum size at which the black hole ceases to radiate. The stable black hole is characterized by a positive heat capacity within the range $r_{min}<r_{H}<r_{crt}$, indicative of a smaller black hole. Conversely, for $r_{H}>r_{crt}$ the system becomes unstable, exhibiting a negative heat capacity associated with a larger, more massive black hole. Additionally, the stable range $r_{min}<r_{H}<r_{crt}$ where the heat capacity remains positive expands as $a$ increases. This observation suggests that smaller values of $a$ result in a quicker cessation of radiation and evaporation for the black hole.

%Fig.\ref{fig:hc} with a PFDM parameter of $\alpha=-0.8$, we see that the heat capacity equals zero at $r_{H}=r_{min}$ which represents the minimum size where the black hole stop radiation.  The stable black hole exhibits a positive heat capacity for $r_{min}<r_{H}<r_{crt}$, which corresponds to a smaller black hole. Conversely, for $r_{H}>r_{crt}$, the system is unstable with a negative heat capacity, indicating a larger, more massive black hole. Furthermore, the stable range $r_{min}<r_{H}<r_{crt}$, where the heat capacity is positive, expands as $a$ increases. This suggests that for smaller values of $a$, the black hole takes less time to stop radiation and evaporation.

{In comparison to existing literature, such as Chaudhary et al. \cite{Chaudhary2024}, which demonstrates that the CoS induces critical regions where black holes transition between stability and instability under EGUP corrections, our findings further highlight that the CoS, in conjunction with PFDM, significantly modifies the thermodynamic properties and stability thresholds of black holes.}

To further examine the phase transition and thermodynamic stability, we now focus on the Gibbs free energy, defined as
\begin{equation}
G=M-TS.\label{28}
\end{equation}
By substituting Eqs. (\ref{massf}), (\ref{temp}), and (\ref{entf}) into Eq. (\ref{28}), we get
\begin{equation}
G=\frac{r_{H}}{4}\left( 1-a-\frac{\alpha }{r_{H}}+\frac{2\alpha }{r_{H}}\ln 
\frac{r_{H}}{\left\vert \alpha \right\vert }\right) .
\end{equation}

\begin{figure}[H]
\centering
\includegraphics[scale=0.7]{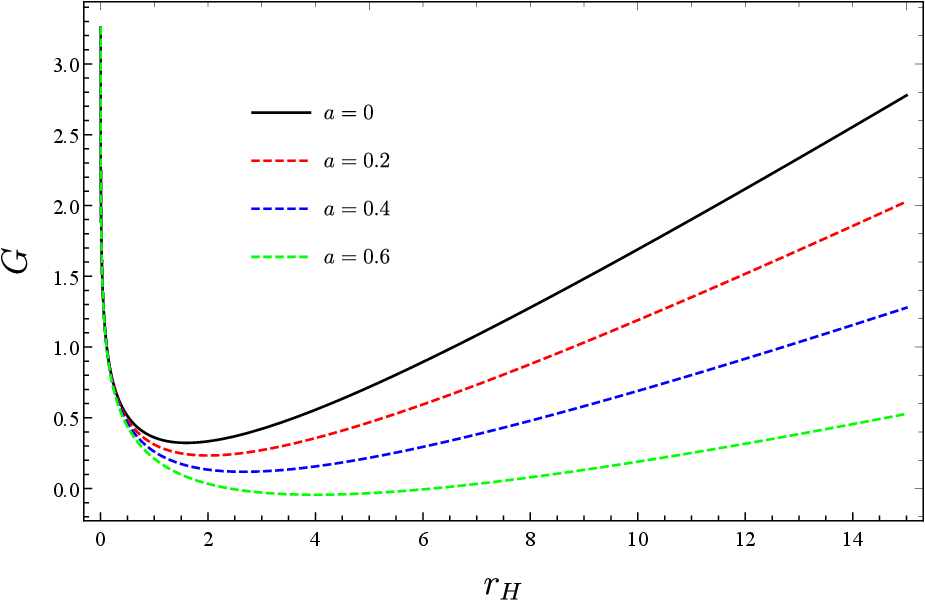}
\caption{Gibbs free energy function versus event horizon with different values of $a$ for $\alpha=-0.8$.}
\label{fig:g}
\end{figure}

\noindent In Fig. \ref{fig:g}, we illustrate the variation of the free energy with respect to $r_{H}$.

\noindent We observe that the Gibbs free energy reaches a minimum at 
\begin{equation}
 r_{H}=-\frac{2\alpha}{1-a}{=r_{crt},}   
\end{equation}
and the value of this horizon shifts with a decrease in the CoS and/or the PFDM parameter, corresponding to a stable region.

%\newpage
\section{Geodesics structure and Shadow} \label{sec4} 
The geodesic equations, along with their corresponding constraint equations, are provided in \cite{Hartle} as follows:
\begin{eqnarray}
\ddot{x}^{\mu }+\Gamma _{\nu \sigma }^{\mu }\dot{x}^{\nu }\dot{x}^{\sigma
}&=&0, \\
g_{\nu \sigma }\dot{x}^{\nu }\dot{x}^{\sigma }&=&-\eta .
\end{eqnarray}%
where $\eta =0$ for null geodesics and $\eta =1$ for timelike geodesics. In this context, the dot indicates differentiation with respect to the affine parameter $s$,  while $x$ refers to the spacetime coordinates. For a given spherically symmetric black hole metric, the geodesic equations take the following form:
\begin{equation}
\ddot{t}+\frac{f^{\prime }\left( r\right) }{f\left( r\right) }\dot{r}\dot{t}=0,  \label{time}
\end{equation}
\begin{equation}
\ddot{r}+\frac{f\left( r\right)}{2} \left( f^{\prime }\left( r\right) \dot{t}%
^{2}+\frac{\dot{r}}{f^{\prime }\left( r\right) }-2r\dot{\theta}^{2}-2r\sin
^{2}\theta \text{ }\dot{\phi}^{2}\right) =0,  \label{radial}
\end{equation}%
\begin{equation}
\ddot{\theta}+\frac{2}{r}\dot{r}\dot{\theta}-\cos \theta \sin \theta \text{ }%
\dot{\phi}^{2}=0,  \label{teta}
\end{equation}%
\begin{equation}
\ddot{\phi}+\frac{2}{r}\dot{r}\dot{\phi}+2\cot \theta \text{ }\dot{\theta}%
\dot{\phi}=0.  \label{phiii}
\end{equation}%
Here, the prime symbol is used to indicate differentiation with respect to the spatial coordinate $r$. After some algebraic manipulation, the constraint for null and timelike geodesics is found to be
\begin{equation}
\left( 1-a-\frac{2M}{r}+\frac{\alpha }{r}\ln \frac{r}{\left\vert \alpha \right\vert }\right) \dot{t}^{2}-\frac{\dot{r}^{2}}{\left( 1-a-\frac{2M}{r}+ \frac{\alpha }{r}\ln \frac{r}{\left\vert \alpha \right\vert }\right) }-r^{2}\left( \dot{\theta}^{2}+\sin ^{2}\theta \dot{\phi}^{2}\right) =-\eta .
\label{cons}
\end{equation}%
These equations, along with Eqs. (\ref{time}) - (\ref{cons}), can be utilized to analyze the behavior of geodesic equations on the equatorial plane, as is done for the SBH in the absence of PFDM and CoS. By setting $\theta =\pi /2$, which implies $\dot{\theta}=\ddot{\theta}=0$, the geodesics equations simplify and one can integrate Eqs. (\ref{time}) and (\ref{phiii}), resulting in:
\begin{eqnarray}
\dot{t}&=&\frac{E}{\left( 1-a-\frac{2c}{r}+\frac{\alpha }{r}\ln \frac{r}{%
\left\vert \alpha \right\vert }\right) },  \label{consE} \\
\dot{\phi}&=&\frac{L}{r^{2}}. \label{consL}
\end{eqnarray}%
Here, the integration constants $E$ and $L$ refer to conserved quantities of the test particles: $E$ corresponds to the total energy, and $L$ represents the angular momentum. Substituting Eqs. (\ref{consE}) and (\ref{consL}) into the constraint equation given in Eq. (\ref{cons}), the energy conservation equation for null geodesics becomes:%
\begin{equation}
\dot{r}^{2}=E^{2}-V_{\mathrm{eff}}\left( r\right) ,  \label{vef}
\end{equation}
where the effective potential is found to be  
\begin{equation}
V_{\rm{eff}}\left( r\right) =\left( 1-a-\frac{2M}{r}+\frac{\alpha }{r}\ln \frac{r}{\left\vert \alpha \right\vert }\right) \left( \eta +\frac{L^{2}}{r^{2}}\right).   \label{potevv}
\end{equation}%
From Eq. (\ref{potevv}), one can analyze both time-like and null geodesics for different values of PFDM and CoS.

\subsection{Timelike geodesics}
The effective potential plays a critical role in analyzing the trajectories of test particles, enabling us to describe their trajectories without explicitly relying on the equations of motion. The extrema of the effective potential delineates stable and unstable circular orbits, with maximum values indicating unstable orbits and minimum values indicating stable ones. For timelike trajectories, where $\eta =1$, the effective potential for particles  can be expressed as
\begin{equation}
V_{\rm{eff}}\left( r\right) =\left( 1-\frac{2M}{r}\right) +\frac{L^{2}}{%
r^{2}}-\frac{2ML^{2}}{r^{3}}-a\left( 1+\frac{L^{2}}{r^{2}}\right) +\frac{
\alpha }{r}\left( 1+\frac{L^{2}}{r^{2}}\right)\ln \frac{r}{\left\vert \alpha \right\vert } .
\end{equation}%
Here, the first component refers to the Newtonian gravitational potential, while the second term accounts for a repulsive centrifugal potential. The third term introduces a relativistic correction arising from general relativity, with a magnitude proportional to a specific factor $1/r^{3}$. Lastly, the fourth and fifth terms stem from the presence of a CoS and PFDM, respectively.

\begin{figure}[htb!]
\begin{minipage}[t]{0.5\textwidth}
        \centering
        \includegraphics[width=\textwidth]{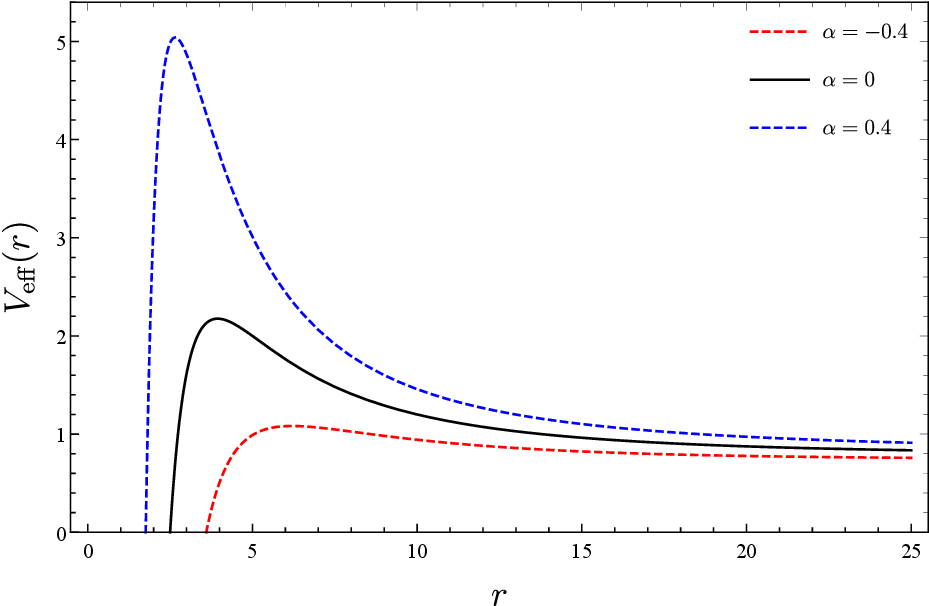}
                \subcaption{$a=0.2$}
        \label{fig:va}
\end{minipage}%$
\begin{minipage}[t]{0.5\textwidth}
        \centering
        \includegraphics[width=\textwidth]{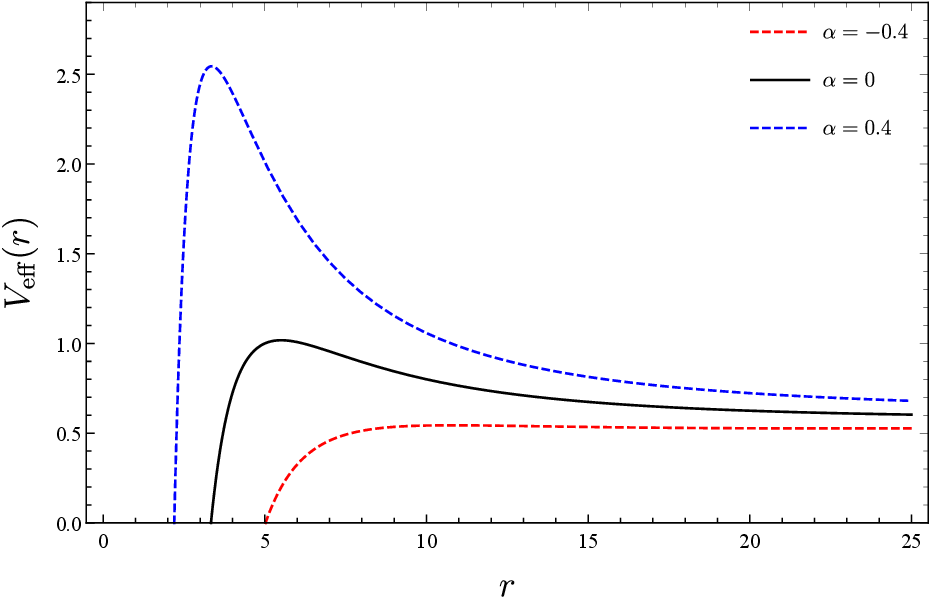}
                \subcaption{$a=0.4$}
        \label{fig:vb}
\end{minipage}
\caption{The behaviors of timelike geodesic effective potential with different values of parameter PFDM $\alpha$, with $L=10$.}
\label{fig:poteff}
\end{figure}

We note that the effective potential is governed by several parameters, including the mass $M$, angular momentum $L$, the CoS parameter $a$, and the PFDM parameter $\alpha $. By fixing two of these parameters $M=1,$ and $L=10$, in Fig. \ref{fig:poteff}. we present the effective potential curves for different values of $\alpha$ in two distinct cases:  $a=0.2,$ and $0.4$.

\noindent The figures demonstrate that as $\alpha$ increases, the maximum peak of the effective potential rises. Moreover, for a fixed value of $\alpha$, increasing the string cloud parameter $a$ causes the overall height of the effective potential to decrease, with the curves shifting downward.

\begin{figure}[H]
\centering
\includegraphics[scale=0.7]{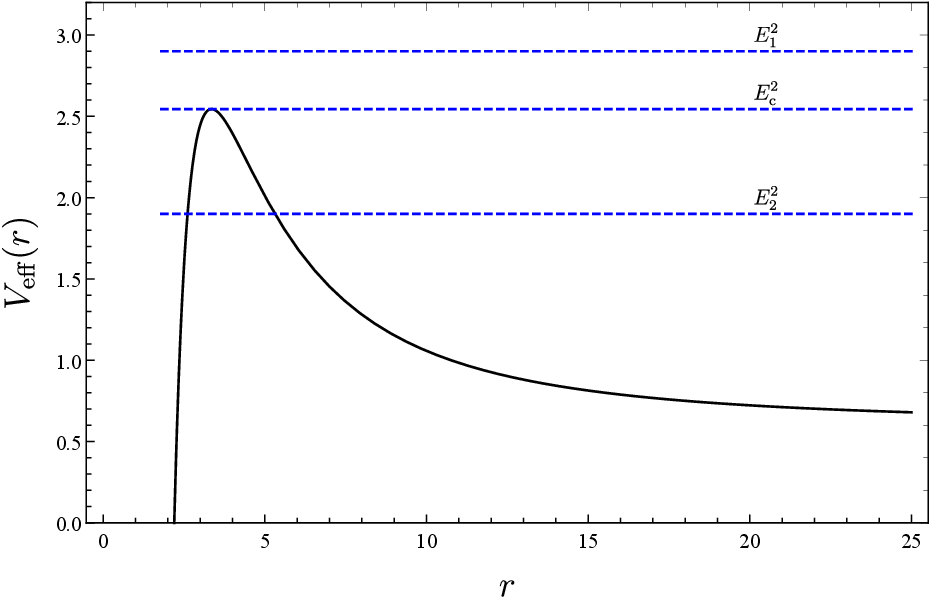}
\caption{Effective potential for a unit mass of black hole, angular momentum value of $L=10$. }
\label{fig:vc}
\end{figure}

Now, let us focus on Eq. (\ref{vef}), which describes the relationship between the energy and the effective potential, where these energy levels govern the motion of particles. We then consider three specific energy levels: $E_{1},E_{2}$ and $E_{c}=E,$ with $E_{c}^{2}-V_{\mathrm{eff}}\left( r\right)=0$ as shown in Fig. \ref{fig:vc}.

%\newpage

\begin{figure}[H]
\centering
\begin{minipage}[t]{0.4\textwidth}
        \centering
        \includegraphics[width=\textwidth]{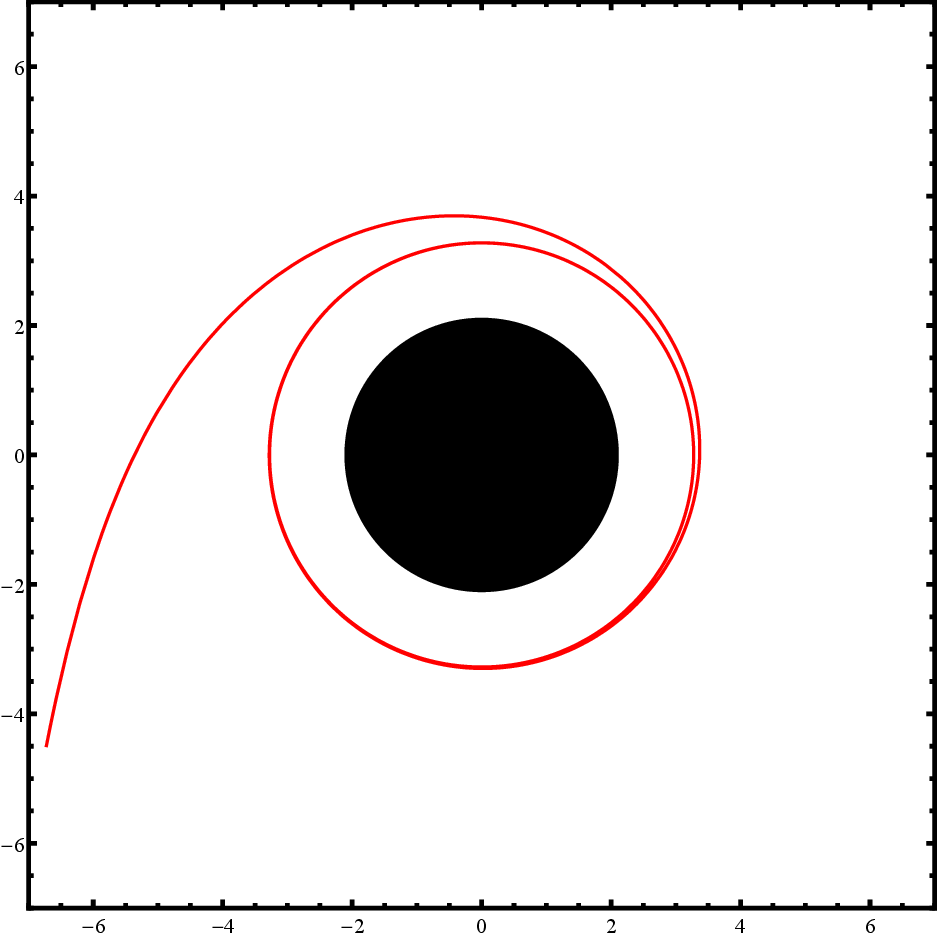}
                \subcaption{$\alpha=0.1$}
        \label{fig:ra}
\end{minipage}%$
\begin{minipage}[t]{0.4\textwidth}
        \centering
        \includegraphics[width=\textwidth]{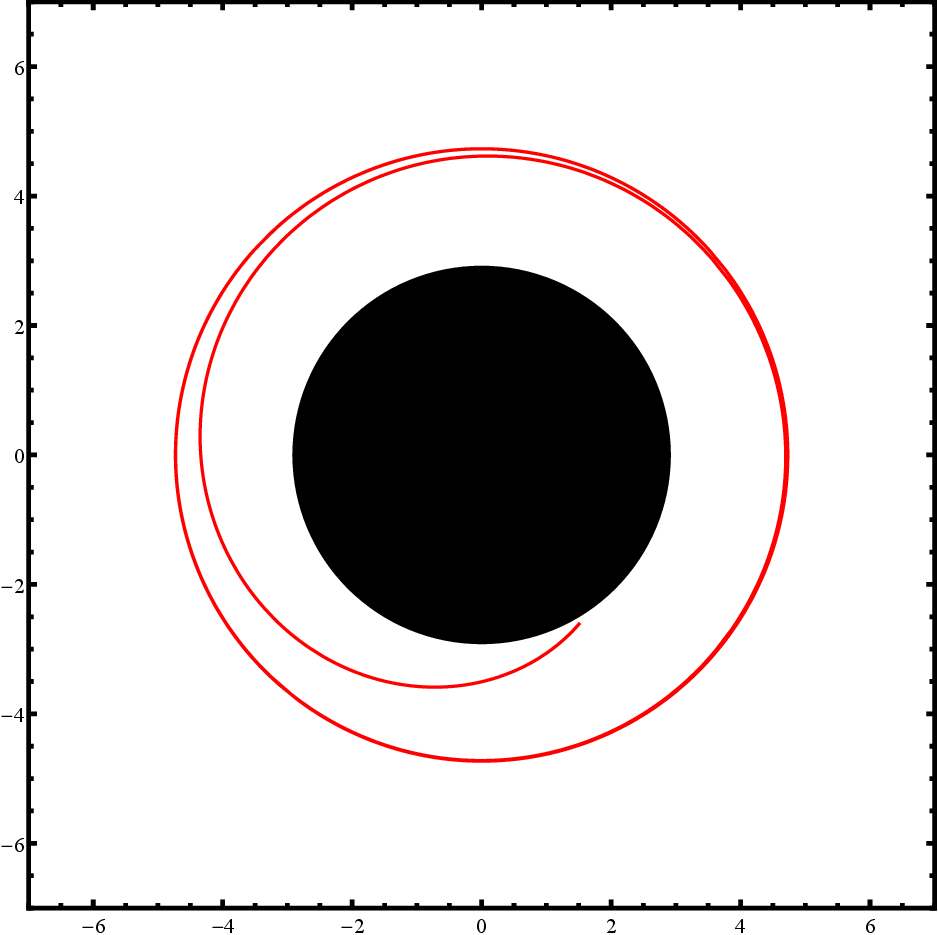}
                \subcaption{$\alpha=-0.1$}
        \label{fig:rb}
\end{minipage}
\caption{Plots of the timelike unstable circular orbits for $L=10$, $M=1$, $a=0.2$.}
\label{fig:rr}
\end{figure}

For these distinct constant energy values, the permitted orbits are as follows:
\begin{itemize}

\item If $E<E_{c}$, the particles can travel from infinity to a minimum distance before returning to infinity. In this scenario, the particles undergo only deflection. The other permitted orbits correspond to photons moving on the opposite side of the potential barrier, ultimately leading them toward the singularity.

\item If $E=E_{c}$ the particles can orbit in an unstable circular trajectory and the orbit radius is influenced by the parameters  $a,$ $M,$ $L,$ and $\alpha$.

\item If $E>E_{c}$, the particles approach from infinity and move partially around the central mass from a very large distance before plunging into the singularity. Thus, the trajectory of the particles does not constitute a stable orbit around the black hole; instead, it ultimately leads to their absorption by the black hole. We demonstrate this scenario in  Fig. \ref{fig:rr}.

\end{itemize}

%\newpage

We will now examine circular orbits and their stability conditions to understand the influence of PFDM and CoS on this type of orbit. A circular orbit is established when the following two conditions are satisfied:
\begin{equation}
\dot{r}=0\Longrightarrow E^{2}-V_{\mathrm{eff}}\left( r\right) =0, \label{32}
\end{equation}
and 
\begin{equation}
\ddot{r}=0\Longrightarrow \left. \frac{d}{dr}V_{\rm{eff}}\left( r\right)
\right\vert _{}=0.  \label{33}
\end{equation}
By solving the above equations, we can obtain the radii of circular orbits for specified values of angular momentum
\begin{equation}
L^{2}=r^{3}\frac{f^{\prime }\left( r\right) }{2f\left( r\right) -rf^{\prime
}\left( r\right) },\label{46}
\end{equation}%
and total energy
\begin{equation}
E^{2}=\frac{2f\left( r\right) ^{2}}{2f\left( r\right) -rf^{\prime }\left(
r\right) }.
\end{equation}
In astrophysics, the innermost stable circular orbit (ISCO) plays a vital role in characterizing the dynamics of a test body in the gravitational field of a compact object. This orbit is described by the convergence of the maximum and minimum points of the effective potential. To accurately determine the ISCO's radius, an additional condition is required: the second derivative of the effective potential with respect to the radius must be zero:
{
\begin{equation}
\frac{d^{2}}{dr^{2}}V_{\mathrm{eff}}\left( r\right) =f^{\prime \prime }\left( r\right) \left( 1+\frac{L^{2}}{r^{2}}\right)
-f^{\prime }\left( r\right) \frac{4L^{2}}{r^{3}}+f\left( r\right) \frac{%
6L^{2}}{r^{4}}=0.  \label{isc}
\end{equation}
Now, by employing Eq. (\ref{46}) along with this condition, we obtain:
\begin{equation}
2rf^{\prime \prime}\left( r\right) f\left( r\right) +6f\left( r\right) f^{\prime }\left(
r\right) -4rf^{\prime }\left( r\right) ^{2}=0.
\end{equation}
}
Before concluding this subsection, we conduct a numerical investigation of the parameters $r_{\rm{ISCO}}$, $L_{\rm{ISCO}}$ and $E_{\rm{ISCO}}$. The results of our calculations for various scenarios are summarized in Table \ref{tablisco} for convenient reference and comparison.
%\begin{table}[tbh]
%\centering{\normalsize \centering%
%\begin{tabular}{l|ll|ll|ll}
%\hline\hline
%\rowcolor{lightgray}\multirow{3}{*}{} & \multicolumn{3}{l|}{$\alpha %=0.1$} & 
%\multicolumn{3}{l|}{$\alpha =0.3$} \\ \hline
%\rowcolor{lightgray}$a$ & $r_{\rm{ISCO}}$ & $L_{\rm{ISCO}}$ & %$E_{\rm{ISCO}}$ & $r_{\rm{ISCO}}$
%& $L_{\rm{ISCO}}$ & $E_{\rm{ISCO}}$ \\ \hline
%$0$ & 5.22562 & 2.89896 & 0.951969 & 4.84986 & 2.41591 & 0.973802 \\ 
%$0.2$ & 6.45311 & 3.57809 & 0.851569 & 5.85946 & 2.89992 & 0.872041 \\ 
%$0.4$ & 8.46861 & 4.69245 & 0.737597 & 7.46782 & 3.66196 & 0.756472 \\ 
%$0.6$ & 12.4167 & 6.87326 & 0.602386 & 10.4855 & 5.06536 & 0.619279 \\ 
%\hline\hline
%\end{tabular}
%\label{tablaa} }
%\caption{•}
%\end{table}
%\begin{table}[tbh]
%\centering{\normalsize \centering%
%\begin{tabular}{l|ll|ll|ll}
%\hline\hline
%\rowcolor{lightgray}\multirow{3}{*}{} & \multicolumn{3}{l|}{$\alpha %=-0.1$}
%& \multicolumn{3}{l|}{$\alpha =-0.3$} \\ \hline
%\rowcolor{lightgray}$a$ & $r_{ISCO}$ & $L_{ISCO}$ & $E_{ISCO}$ & %$r_{ISCO}$
%& $L_{ISCO}$ & $E_{ISCO}$ \\ \hline
%$0$ & 6.87751 & 4.0847 & 0.936026 & 7.7918 & 4.83513 & 0.925127 \\ 
%$0.2$ & 8.68426 & 5.15634 & 0.837268 & 10.0209 & 6.2065 & 0.827892 \\ 
%$0.4$ & 11.7291 & 6.96182 & 0.725161 & 13.8429 & 8.5538 & 0.717428 \\ 
%$0.6$ & 17.9108 & 10.6259 & 0.592165 & 21.7779 & 13.4167 & 0.586256 \\ 
%\hline\hline
%\end{tabular}
%\label{tablba} }
%\caption{.}
%\end{table}
%%%%%%%%%%%%%%%%%%%%%%%%%%%%%%%%%%%%%%%%%%%%%%%%%%%%%%%%%%%%%%%%%%%%%%%
\begin{table}[tbh]
\centering{\normalsize \centering%
\begin{tabular}{l|lll|lll}
\hline\hline
%\rowcolor{lightgray}\multirow{3}{*}& \multicolumn{3}{l|}\rowcolor{lightgray}\multirow{3}{*}& \multicolumn{3}{l|} & \multicolumn{3}{l|}{$\alpha =0.3$} \\ \hline

\rowcolor{lightgray}$a$ & $r_{\rm{ISCO}}$ & $L_{\rm{ISCO}}$ & $E_{\rm{ISCO}}$ & $r_{\rm{ISCO}}$& $L_{\rm{ISCO}}$ & $E_{\rm{ISCO}}$ \\ \hline

$0$   & 5.22562 & 2.89896 & 0.951969 & 4.84986 & 2.41591 & 0.973802 \\ 
$0.2$ & 6.45311 & 3.57809 & 0.851569 & 5.85946 & 2.89992 & 0.872041 \\ 
$0.4$ & 8.46861 & 4.69245 & 0.737597 & 7.46782 & 3.66196 & 0.756472 \\ 
$0.6$ & 12.4167 & 6.87326 & 0.602386 & 10.4855 & 5.06536 & 0.619279 \\ 
\rowcolor{lightgray}\multirow{3}{*}{} & \multicolumn{3}{l|}{$\alpha =-0.1$}& \multicolumn{3}{l|}{$\alpha =-0.3$} \\ \hline

$0$   & 6.87751 & 4.0847 & 0.936026 & 7.7918 & 4.83513 & 0.925127 \\ 
$0.2$ & 8.68426 & 5.15634 & 0.837268 & 10.0209 & 6.2065 & 0.827892 \\ 
$0.4$ & 11.7291 & 6.96182 & 0.725161 & 13.8429 & 8.5538 & 0.717428 \\ 
$0.6$ & 17.9108 & 10.6259 & 0.592165 & 21.7779 & 13.4167 & 0.586256 \\ 
\hline\hline
\end{tabular}
\caption{The numerical values of the ISCO parameters $r_{\rm{ISCO}}$, $L_{\rm{ISCO}}$ and $E_{\rm{ISCO}}$ for test particles are tabulated for various possible scenarios.}
\label{tablisco} }
\end{table}

\subsection{Null geodesics}
\begin{figure}[H]
\begin{minipage}[t]{0.5\textwidth}
        \centering
        \includegraphics[width=\textwidth]{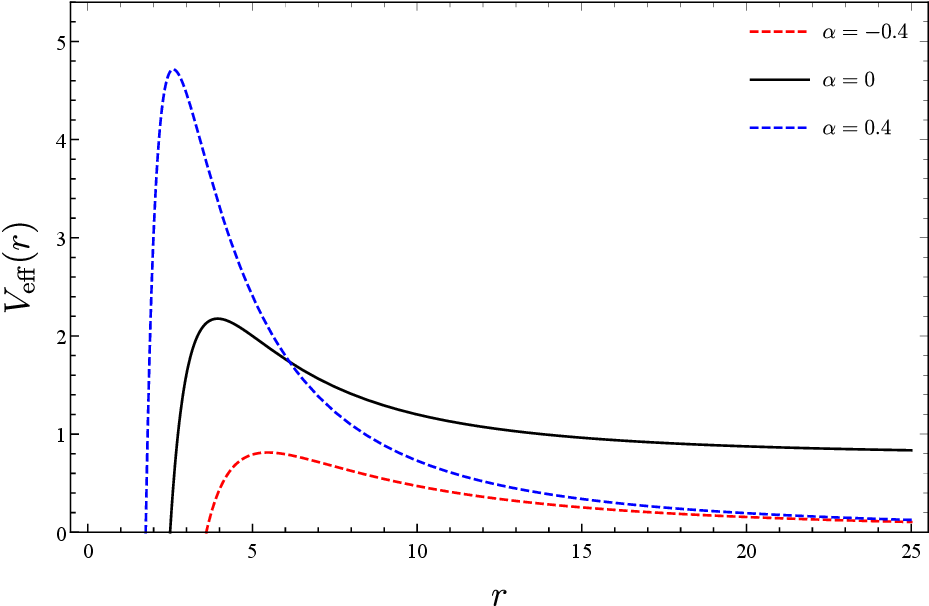}
                \subcaption{$a=0.2$}
        \label{fig:vla}
\end{minipage}%$
\begin{minipage}[t]{0.5\textwidth}
        \centering
        \includegraphics[width=\textwidth]{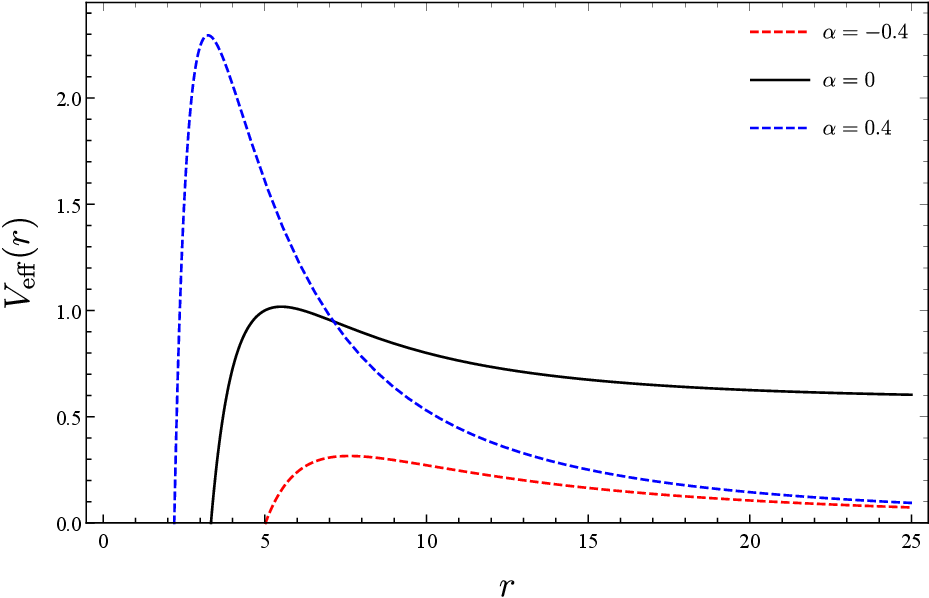}
                \subcaption{$a=0.4$}
        \label{fig:vlb}
\end{minipage}
\caption{Effective potential for null geodesics with different values of the PFDM parameter for $L=10$.}
\label{fig:vefflight}
\end{figure}

For lightlike particles, where $\eta =0$, the effective potential reduces to
\begin{equation}
V_{\mathrm{eff}}\left( r\right) =\left( 1-a-\frac{2M}{r}+\frac{\alpha }{r}%
\ln \frac{r}{\left\vert \alpha \right\vert }\right) \frac{L^{2}}{r^{2}}.
\end{equation}
Fig. \ref{fig:vefflight}  illustrates the effective potential for $a=0.4$ and $0.2$ with various PFDM values.

\noindent The graph shows the effective potential rising to a peak at the photon sphere radius, denoted as 
$r_{\rm{ph}}$, before declining to zero. The behavior of lightlike geodesics is notably simpler than that of timelike geodesics. Each curve for the effective potential exhibits a single energy maximum, which characterizes an unstable circular orbit. As the PFDM increases, the maximum value of the effective potential also increases. Conversely, when the CoS parameter decreases, the peak value of the effective potential rises.

We now examine the radius of the photon sphere $r_{\rm{ph}}$, and the associated instability condition to understand how the presence of PFDM and a string cloud affect it. For this purpose, we consider the scenario of circular orbits, $\dot{r}=0$, where the corresponding effective potential must satisfy the following conditions: 
\begin{eqnarray}
V_{\mathrm{e}}=V_{\rm{eff}}\left( r_{\rm{ph}}\right) -E^{2}&=&0,  \label{cond1} \\
\left.V_{\mathrm{e}}^{\prime}\right\vert _{r=r_{\rm{ph}}}&=&0.
\label{cond2}
\end{eqnarray}
Additionally, the condition for instability requires that: 
\begin{equation}
    V_{e}^{\prime \prime} \big\vert
_{r=r_{\rm{ph}}}<0. \label{thirdcond}
\end{equation}
Using the first condition given in Eq. \eqref{cond1}, we obtain
\begin{equation}
\xi=\sqrt{\frac{r_{\rm{ph}}^{2}}{\left( 1-a-\frac{2M}{r_{\rm{ph}}}+\frac{\alpha }{r_{\rm{ph}}}\ln \frac{r_{\rm{ph}}}{\left\vert \alpha \right\vert }\right) }},
\end{equation}
where $\xi$ is the impact parameter, defined as $\xi=\frac{L}{E}$. Using the second condition, given in Eq. \eqref{cond2}, we find
\begin{equation}
2\left( a-1\right) +\frac{6M}{r_{\rm{ph}}}-\frac{3\alpha }{r_{\rm{ph}}}\ln \frac{r_{\rm{ph}}}{\left\vert \alpha \right\vert}+\frac{\alpha }{r_{\rm{ph}}}=0.  \label{rp}
\end{equation}
In this case, finding an exact analytical solution to Eq. (\ref{rp}) is not feasible. Therefore, we resort to numerical methods to solve it. The introduction of PFDM and the string cloud introduces two new parameters into Eq. (\ref{rp}). We will explore various values for these parameters and employ numerical techniques to determine the radius of the photon sphere. Subsequently, we will compute the corresponding impact parameter $\xi$ and present the results in Table \ref{tCb}.
\begin{table}[H]
\centering
{\normalsize \centering%
\begin{tabular}{l|ll|ll|ll}
\hline\hline
\rowcolor{lightgray}\multirow{2}{*}{} & \multicolumn{2}{l|}{$\alpha =0.1$} & 
\multicolumn{2}{l|}{$\alpha =0.5$} & \multicolumn{2}{l}{$\alpha =0.9$} \\ 
\hline
\rowcolor{lightgray}$a$ & $r_{\rm{ph}}$ & $\xi $ & $r_{\rm{ph}}$ & $\xi $ & $r_{\rm{ph}}$ & $%
\xi $ \\ \hline
$0$ & 2.56341 & 4.35581 & 2.15448 & 3.36190 & 2.22692 & 3.25507 \\ 
$0.2$ & 3.16475 & 6.01096 & 2.53910 & 4.40464 & 2.55304 & 4.11901 \\ 
$0.4$ & 4.15180 & 9.10280 & 3.12567 & 6.21022 & 3.02350 & 5.52730 \\ 
$0.6$ & 6.08438 & 16.3306 & 4.15484 & 9.97632 & 3.78084 & 8.19831 \\ 
$0.8$ & 11.6797 & 44.2970 & 6.58355 & 21.7074 & 5.29200 & 15.0674 \\ 
\rowcolor{lightgray}\multirow{2}{*}{} & \multicolumn{2}{l|}{$\alpha =-0.1$}
& \multicolumn{2}{l|}{$\alpha =-0.5$} & \multicolumn{2}{l}{$\alpha =-0.9$}\\ \hline
$0$ & 3.48255 & 6.12047 & 4.37716 & 8.05550 & 4.81372 & 9.24671 \\ 
$0.2$ & 4.3969 & 8.63824 & 5.72275 & 11.7418 & 6.53234 & 13.9036 \\ 
$0.4$ & 5.93764 & 13.4673 & 8.05812 & 19.0293 & 9.56867 & 23.3003 \\ 
$0.6$ & 9.06513 & 25.1755 & 12.9812 & 37.3966 & 13.1115 & 51.9985 \\ 
$0.8$ & 18.6722 & 73.3053 & 28.9732 & 117.392 & 38.0181 & 156.816 \\ 
\hline\hline
\end{tabular}
\label{tablCb} }
\caption{The photon sphere radius and the impact parameter for different values of PFDM and CoS with $M=1$.}
\label{tCb}
\end{table}

\begin{figure}[H]
\begin{minipage}[t]{0.5\textwidth}
        \centering
        \includegraphics[width=\textwidth]{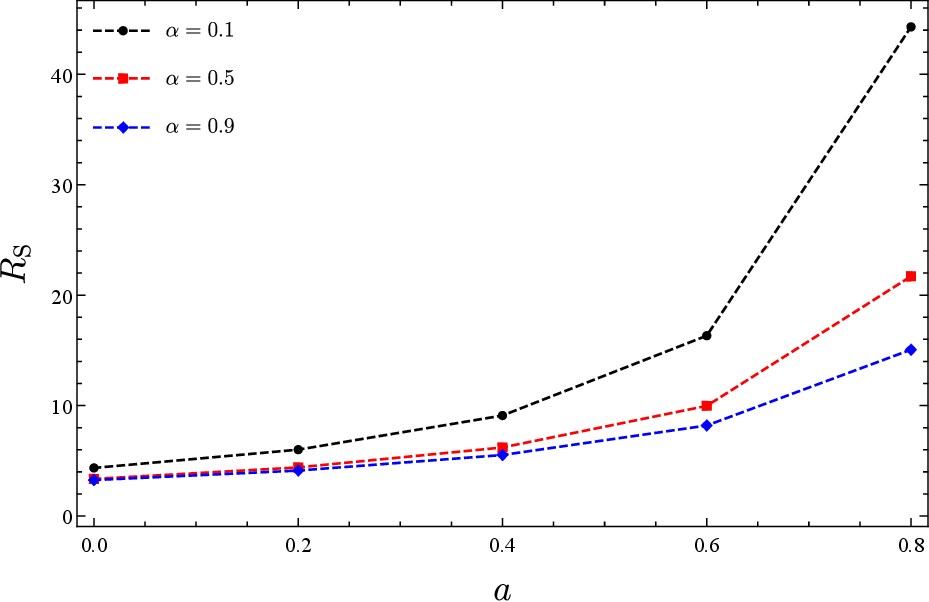}
         \label{fig:sha}
\end{minipage}%$
\begin{minipage}[t]{0.5\textwidth}
        \centering
        \includegraphics[width=\textwidth]{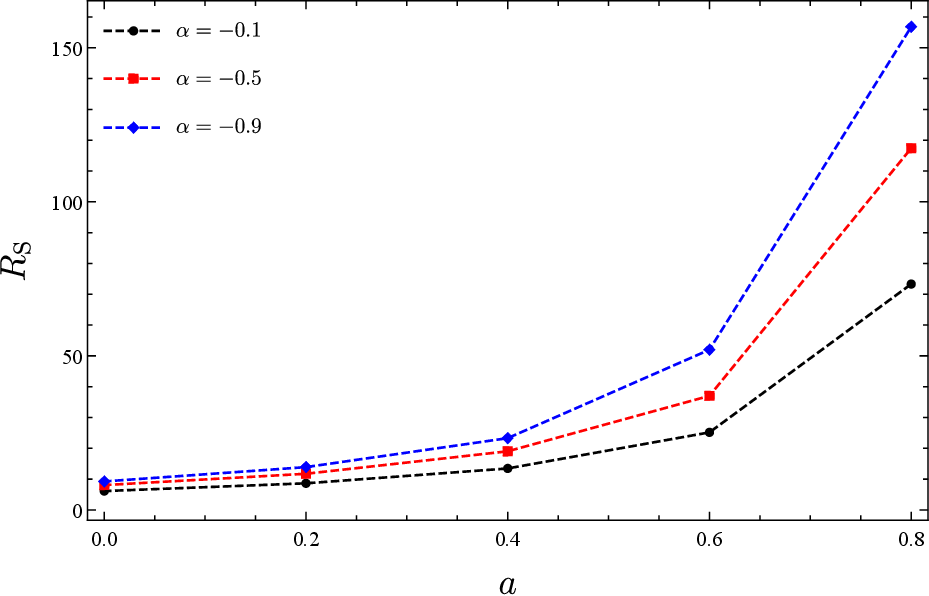}
        \label{fig:shb}
\end{minipage}
\caption{Black hole shadow radius as a function of CoS for various PFDM parameters.}
\label{fig:shadow}
\end{figure}

Let us examine a light ray emitted by a stationary observer located at $r_{\rm{O}}$, traveling into the past. The angle $\Theta $ between such a light ray and the radial direction is given by
\begin{equation}
\sin ^{2}\Theta =\frac{f\left( r_{\rm{O}}\right) R^{2}}{r_{\rm{O}%
}^{2}f\left( R\right) }.  \label{22}
\end{equation}%
Within this framework, the angular radius of the black hole shadow can be
determined by taking the limit $R\rightarrow r_{\rm{ph}}$ in Eq. (\ref{22})
\begin{equation}
\sin \Theta _{\rm{S}}=\frac{r_{\rm{ph}}}{r_{\rm{O}}}\sqrt{\frac{f\left(r_{\rm{O}}\right) }{f\left( r_{\rm{ph}}\right) }}.
\end{equation}%
When $r_{\rm{O}}=r_{\rm{ph}}$, the angle $\Theta _{\rm{S}}=\pi /2$, indicating that the shadow covers exactly half of the observer's sky. For a static observer located at a large distance, $r_{\rm{O}}\rightarrow \infty$, the shadow radius of the black hole is found to be
\begin{equation}
R_{\rm{S}}={\frac{r_{\rm{ph}}}{\sqrt{f\left( r_{\rm{ph}}\right) }}}=\xi.
\end{equation}
With the black hole mass fixed at $M=1$, the shadow radii are illustrated in Fig. \ref{fig:shadow}.

\begin{figure}[H]
 \centering
        \includegraphics[scale=0.7]{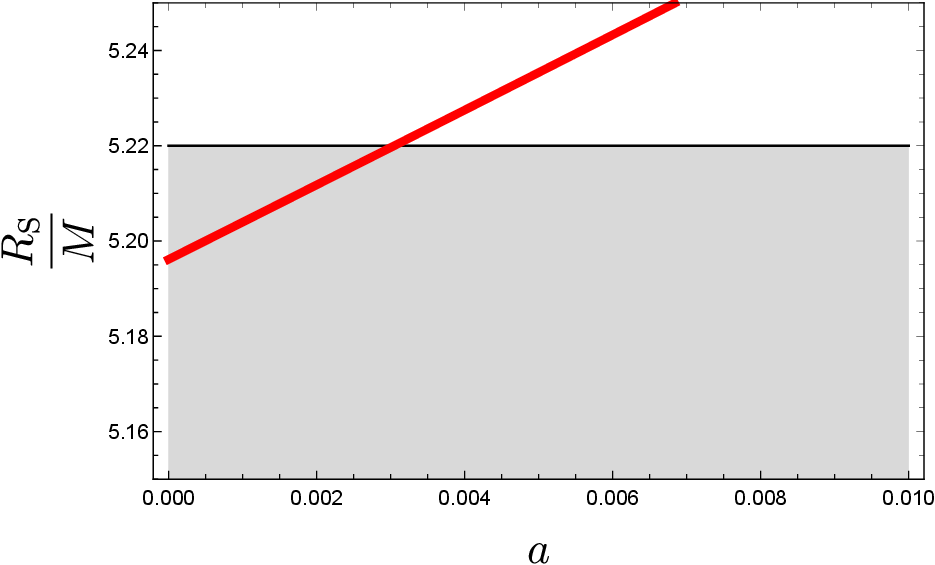}
        \caption{Shadow radius $\frac{R_{\rm{S}}}{M}$ of the SBH surrounded by CoS as a function of the string parameter.}
\label{fig:sha1dow}
\end{figure}

Using the equation above, we analyze the constraints on the parameter $a$ based on the EHT data, as outlined in Table \ref{tabsrg}.
\begin{table}[htb!]
\centering%
\begin{tabular}{l|ll|l}
\hline\hline
& Mass ($\rm M_{\odot }$) & Angular diameter ($\rm \mu as$) & Distance($\rm kpc$) \\ 
\hline
Sgr. A$^{\ast }$ & $\left( 4.13\pm 0.013\right) \times 10^{6}$ & $48.7\pm 7$
& $8.277\pm 0.033$ \\ 
\hline
\end{tabular}%
\caption{Observational measurements of Sgr. A$^{\ast }$  black hole.}
\label{tabsrg}
\end{table}
To simplify the calculation,  we use the metric of the SBH surrounded by a CoS in the absence of PFDM. In this scenario, the radius of the photon sphere and the radius of the black hole shadow are given by, respectively:
\begin{eqnarray}
r_{\rm{ph}}&=&\frac{3M}{\left( 1-a\right) }, \\
\frac{R_{\rm{S}}}{M}&=&\frac{3\sqrt{3}}{\left( 1-a\right) ^{3/2}}.  \label{R30}
\end{eqnarray}
{ It is worth noting that our results indicate an inverse proportionality between the photon sphere radius and $(1-a)$, a behavior also observed in the literature. In particular, in \cite{Badawi2024}  Al-Badawi et al. demonstrated that both the photon sphere radius and the impact factor, influenced by the presence of the CoS, exhibit the same inverse scaling with $(1-a)$.

Then, we} follow the bounds provided in \cite{Sunny}. For Sgr A$^{\ast }$ the permitted range is
\begin{equation}
    4.55\leq \frac{R_{\rm{S}}}{M}\leq 5.22.
\end{equation}
\noindent Our result is presented in Fig. \ref{fig:sha1dow}, which includes the confidence interval. 

%\begin{figure}[H]
%\begin{minipage}[t]{0.5\textwidth}
%        \centering
%        \includegraphics[width=\textwidth]{srg.eps}
%        \subcaption{ Sgr. A$^{\ast }$.}
 %       
%\end{minipage}%$
%\begin{minipage}[t]{0.5\textwidth}
%        \centering
%        \includegraphics[width=\textwidth]{m87.eps}
%        \subcaption{ M87$^{\ast }$.}
%    \end{minipage}
%\caption{Shadow radius $\frac{R_{\rm{S}}}{M}$ of the SBH %surrounded by CoS as a function of the string parameter.}
%\label{fig:sha1dow}
%\end{figure}

\noindent While the overall shape of the curve stays consistent because of Eq. (\ref{R30}), we identify distinct upper bounds for $a$, which are presented in Table \ref{datasig}. 
%Although the general shape of the curve remains similar due to the use of Eq. (\ref{R30}), we observe distinct upper bounds for $a$. These are provided in Table \ref{datasig}. While the constraints on $a$ for Sgr A$^{\ast }$ and M87$^{\ast }$ differ, the bounds are relatively close when compared. 
\begin{table}[htb!]
\centering%
\begin{tabular}{l|ll}
\hline\hline
& 1$\mathbb{\sigma }$ upper & 1$\mathbb{\sigma }$ lower \\ \hline
Sgr. A$^{\ast }$ & $0.003$ & $0$ \\ 
\hline
\end{tabular}%
\caption{Values of $a$ based on the constraints imposed by the EHT data on the shadow radius.}
\label{datasig}
\end{table}

\section{Quasinormal modes} \label{sec5}
The investigation of black hole quasinormal modes (QNMs) is a long-established and thoroughly studied area. The distinct frequencies of QNMs provide insights into the black hole's characteristics as well as the types of radiative emissions it can produce. Originally, these frequencies were computed using purely numerical methods. This involved selecting a complex frequency, solving the corresponding differential equation through numerical integration, and then verifying if the solution met the necessary boundary conditions. To our knowledge, researchers have developed a variety of methods to study QNMs of black holes. These approaches include numerical Integration Methods \cite{Berti}, WKB Approximation \cite{Schutz, Iyer, Konoplya}, Leaver's Continued Fraction Method \cite{Leaver}, Pöschl-Teller Potential Approximation \cite{Valeria}, Asymptotic Iteration Method \cite{Cho}, Time-domain Integration \cite{Carsten}. Additionally, it has been shown that the characteristics of QNMs can be related to the properties of photons trapped in unstable circular orbits around black holes \cite{Witek, Ivan, Jusufi}. In this section, we derive QNMs using two methods: the sixth-order WKB approximation method and the shadow radius method.

\subsection{WKB method}
In curved spacetime, the dynamics of a massless scalar field are governed by the following equation:
\begin{equation}
\frac{1}{\sqrt{-g}}\partial _{\mu }\left( \sqrt{-g}g^{\mu \nu }\partial
_{\nu }\psi \right) =0.  \label{klei}
\end{equation}
To solve this, we adopt the ansatz 
\begin{equation}
\psi \left( t,r,\theta ,\varphi \right) =e^{-i\omega t}\frac{\Psi _{\omega
,{\ell}}\left( r\right) }{r}Y_{{\ell},\mu }\left( \theta ,\varphi \right) ,
\label{dec}
\end{equation}%
where $\omega$ represents the frequency, {$\ell$ is the angular quantum number}, and $Y_{{\ell},\mu }\left( \theta ,\varphi\right)$ are the spherical harmonics. Substituting this decomposition into Eq. (\ref{klei}) transforms it into a Schr\"{o}dinger-like equation:

\begin{equation}
\frac{d^2}{dr^{\ast ^2 }}\Psi \left( r^{\ast }\right) -\left( \omega ^{2}-%
\mathcal{V}\left( r\right) \right) \Psi \left( r^{\ast }\right) =0,
\label{eqto}
\end{equation}
where the tortoise coordinate $r^{\ast }$ is related to $r$ via $dr^{\ast }=\frac{dr}{f\left( r\right) }$. The effective potential $\mathcal{V}\left( r\right)$ is given by
\begin{equation}
\mathcal{V}\left( r\right) =\frac{f\left( r\right) }{r} f^{\prime}(r)+\frac{f\left( r\right) {\ell\left( \ell-1\right) } }{r^2}.
\end{equation}
To solve Eq. (\ref{eqto}), it is essential to impose appropriate boundary conditions. In this context, the physically acceptable solutions are those that represent purely ingoing waves near the black hole horizon, expressed as:
\begin{equation}
\Psi \simeq e^{\pm i\omega r^{\ast }},\qquad r^{\ast }\rightarrow
\pm \infty .
\end{equation}
Next, utilizing the WKB approximation, we compute the QNM frequencies using the following relation:
\begin{equation}
i\frac{\left( \omega -V_{0}\right) }{\sqrt{-2V_{0}"}}-\sum_{i=2}^{N}\Lambda
_{i}=n+\frac{1}{2}.  \label{qnmf}
\end{equation}
where $V_{0}$ corresponds to the height of the effective potential, and 
$V_{0}^{\prime \prime}$ denotes the second derivative of the potential with respect to the tortoise coordinate, evaluated at its maximum $r_{0}^{\ast }$, respectively. The term $\Lambda _{i}$ accounts for corrections from higher-order terms in the WKB approximation, while $n=0,1,2,...$, is the overtone number.

Now, we calculate the QNM frequencies of the SBH by applying the 6th-order WKB approximation method for different angular quantum numbers {$\ell$} and overtone numbers $n$ in the context of a massless scalar field. These frequencies are separated into real and imaginary components, where the real part refers to the oscillation frequency and the imaginary part corresponds to the decay rate. We tabulate the QNMs in Table \ref{tab:week3}. 
\begin{table}[tbh]
\centering%
\begin{tabular}{|l|l|l|l|l|l|}
\hline\hline
\multicolumn{2}{|c|}{} &
%\multicolumn{2}{*}{} &
\multicolumn{2}{|c|}{$\alpha =0.3$} &
\multicolumn{2}{|c|}{$\alpha =0.6$}\\ \hline
\rowcolor{lightgray} ${\ell}$ & $n$ & $\omega _{\rm{WKB}}(a=0.2)$ & $%
\omega _{\rm{WKB}}(a=0.5)$ & $\omega _{\rm{WKB}}(a=0.2)$ & $%
\omega _{\rm{WKB}}(a=0.5)$ \\ \hline
1 & 0 & 0.306997 - i 0.101910 & 0.162166 - i 0.0436177 & 0.353318 - i 0.128573 & 0.202572 - i 0.060770 \\ \hline 2 & 0 & 0.510563 - i 0.100880 & 0.273616 - i 0.0433136 & 0.586499 - i 0.126928 & 0.341274 - i 0.0602066 \\ & 1 & 0.488732 - i 0.308314  & 0.265861 - i 0.131531 & 0.555395 - i 0.389586 & 0.328738 - i 0.183452 \\ \hline 3 & 0 & 0.714401 - i 0.100589  & 0.384350 - i 0.0432315 & 0.820241 - i 0.126460 & 0.479200 - i 0.0600535 \\ & 1 & 0.698288 - i 0.304695 & 0.378717 - i 0.130505 & 0.797165 - i 0.383912 & 0.470059 - i 0.181602  \\  & 2 & 0.668314 - i 0.517244 & 0.367940 - i 0.220160 & 0.754681 - i 0.654423 & 0.452703 - i 0.307391 \\ \hline 4 & 0 & 0.918318 - i 0.100469 & 0.494845 - i 0.0431980 & 1.05415 - i 0.126265 & 0.616865 - i 0.0599908 \\ & 1 & 0.905622 - i 0.303186 & 0.490435 - i 0.130084 & 1.03594 - i 0.381548 & 0.609698 - i 0.180844 \\ & 2 & 0.881281 - i 0.511134 & 0.481841 - i 0.218425 & 1.00120 - i 0.644934 & 0.595790 - i 0.304287 \\ & 3 & 0.847393 - i 0.727322  & 0.469526 - i 0.309120 & 0.953375 - i 0.921010  & 0.576018 - i 0.431912 \\ \hline\hline
\end{tabular}%
\caption{QNMs of the SBH surrounded by PFDM and CoS, calculated using the 6th-order WKB approximation method.}
\label{tab:week3}
\end{table}

\newpage 

\noindent We observe that the imaginary components of all frequencies are negative, confirming the stability of the black hole. {This also indicates the damping behavior of the oscillations, which is tied to the dynamical evolution of the QNMs. }

\begin{figure}[H]
\begin{minipage}[t]{0.5\textwidth}
        \centering
        \includegraphics[width=\textwidth]{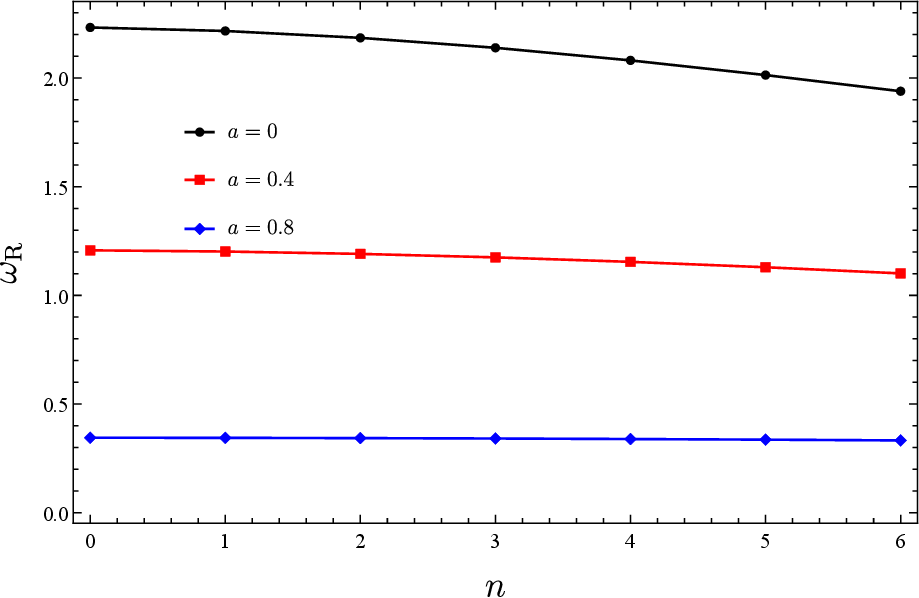}
            \label{fig:qa}
\end{minipage}%$
   \begin{minipage}[t]{0.5\textwidth}
        \centering
        \includegraphics[width=\textwidth]{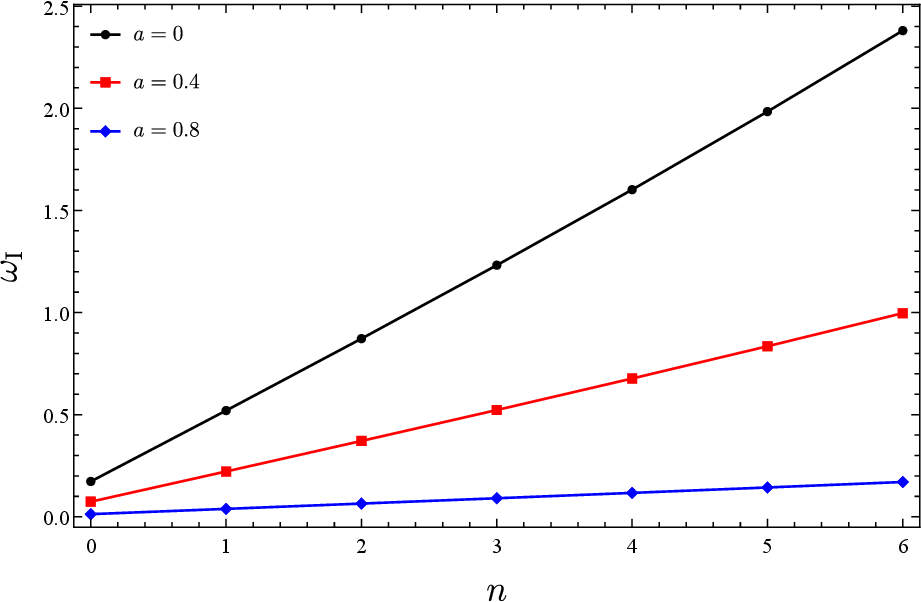}
               \label{fig:qb}
\end{minipage}
\caption{The quasinormal modes of the black holes in scalar field for the state {$\ell=7$}, $\alpha=0.5$ and $M=1$.}
\label{fig:qnm1}
\end{figure}

Fig. \ref{fig:qnm1} illustrates the variation of the QNM frequencies of the SBH as a function of the overtone numbers for different values of the CoS parameter. 

\noindent It is clear that for a fixed $n$, both $\omega_{\rm{R}}$ and $\omega_{\rm{I}}$ decrease as $a$ increases. This suggests that a reduction in $a$ results in an increase in the oscillation frequency and a faster decay rate. { This result is consistent with the findings presented by Al-Badawi et al. in Table 2 of  \cite{Badawi2024}.}

\begin{figure}[H]
\begin{minipage}[t]{0.5\textwidth}
        \centering
        \includegraphics[width=\textwidth]{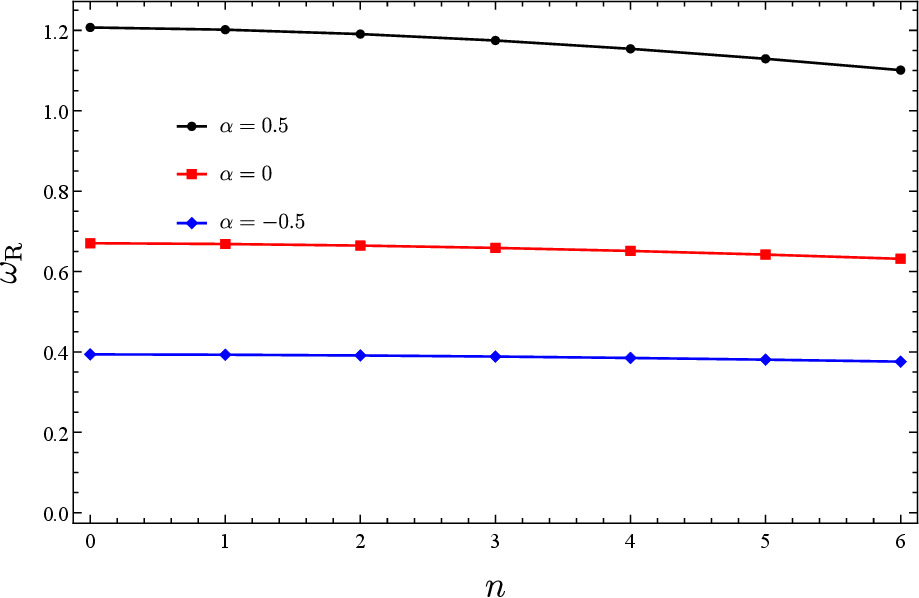}
            \label{fig:qc}
\end{minipage}%$
   \begin{minipage}[t]{0.5\textwidth}
        \centering
        \includegraphics[width=\textwidth]{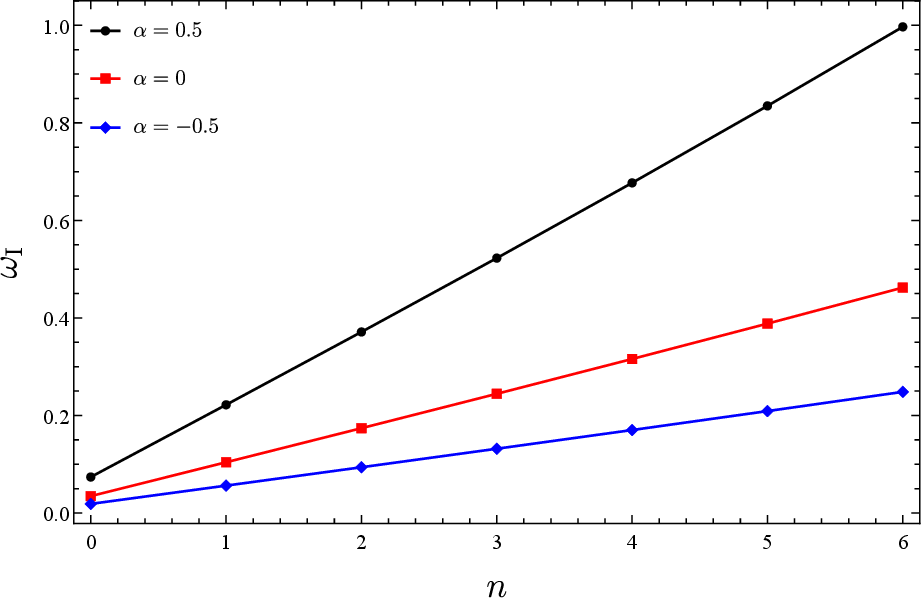}
               \label{fig:qd}
\end{minipage}
\caption{QNMs of the SBH in the scalar field for the state {$\ell=7$}, $a=0.4$ and $M=1$.}
\label{fig:qnm2}
\end{figure}

Next, by setting $a=0.4$ we examine the impact of PFDM on the QNMs, as depicted in Fig. \ref{fig:qnm2}.

\noindent It is apparent that for a fixed overtone number, both components of the frequencies,  $\omega_{\rm{R}}$ and $\omega_{\rm{I}}$, increase with increasing PFDM parameter. This implies that a higher $\alpha$ enhances the oscillation frequency, and decelerates the rate of decay.

\subsection{Connection between shadow and quasinormal modes}

Previous studies \cite{Witek, Ivan, Jusufi} have established that in the eikonal limit, the real component of the QNMs is related to the angular velocity of the final circular null geodesic, whereas the imaginary component is linked to the Lyapunov exponent,
\begin{equation}
\omega _{\rm{S}} =\omega _{\rm{R}}-i\omega _{\rm{I}}=\Omega {\ell}-i\lambda \left(n+\frac{1}{2}\right),\label{omegasha}
\end{equation}%
where
\begin{equation}
\Omega =\frac{\sqrt{f\left( r_{\rm{ph}}\right) }}{r_{\rm{ph}}},
\end{equation}
represents the angular velocity of the photon sphere located at the unstable orbit, and
\begin{equation}
\lambda =\left. \sqrt{-\frac{V^{\prime \prime}\left( r\right) }{2\dot{t}^{2}}}\right\vert_{r=r_{\rm{ph}}}=\sqrt{\frac{\left( 2f\left( r_{\rm{ph}}\right) -r_{\rm{ph}}^{2}f^{\prime \prime }\left( r_{\rm{ph}}\right) \right) f\left( r_{\rm{ph}}\right) }{2r_{\rm{ph}}^{2}}},
\end{equation}%
denotes the Lyapunov exponent. 

Now, we compute quasinormal frequencies for scalar field perturbations across different angular momentum, PFDM and  CoS parameter values, utilizing  Eqs. (\ref{qnmf}) and (\ref{omegasha}).  By comparing the quasinormal frequencies for various values of angular momentum, we tabulate the results in Tables \ref{tab:q1}, \ref{tab:q2}, \ref{tab:q3}, and \ref{tab:q4}.
\begin{table}[H]
\centering{\normalsize \centering%
\begin{tabular}{l|l|l|ll}
\hline\hline
%\rowcolor{lightgray}\multirow{4}{*} & \multicolumn{4}{l}{$\alpha =0.1$ and $a=0.2$} \\ 
\hline

\rowcolor{lightgray}{$\ell$} & $\omega _{\rm{WKB}}$ & $\omega _{\rm{S}}$ & $\Delta _{\rm{R}} \%$ & $\Delta _{\rm{I}} \%$ \\ \hline\hline

$5$ & 0.915073 - i 0.0766499  & 0.83182 - i 0.076572 & 10.009 & 0.10173 \\ 
$7$ & 1.24778 - i 0.0766139   & 1.1645 - i 0.076572  & 7.1516 & 5.4720 $\times 10^{-2}$\\ 
$10$ & 1.74685 - i 0.0765934  & 1.6636 - i 0.076572 & 5.0023 &  2.7948 $\times 10^{-2}$ \\ 
$15$ & 2.57865 - i 0.0765818  & 2.4954 - i 0.076572  & 3.3361 & 1.2798 $\times 10^{-2}$ \\ 
$20$ & 3.41046 - i 0.0765776  & 3.3273 - i 0.076572  & 2.4993 &  7.3134 $\times 10^{-3}$ \\ 
$30$ & 5.07408 - i 0.0765745  & 4.9909 - i 0.076572   & 1.6666 &  3.2649 $\times 10^{-3}$ \\
$40$ & 6.73771 - i 0.0765734 & 6.6545 - i 0.076572  & 1.2504 & 1.8283 $\times 10^{-3}$\\
$50$ & 8.40134 - i 0.0765729 & 8.3182 - i 0.076572  & 0.99950 &  1.1754 $\times 10^{-3}$ \\ 
\hline\hline
\end{tabular}
}
\caption{QNMs calculated by 6th-order WKB approximation method and by the shadow radius, where $a=0.2$ and $\alpha =0.1$.}
\label{tab:q1}
\end{table}

\begin{table}[H]
\centering{\normalsize \centering%
\begin{tabular}{l|l|l|ll}
\hline\hline
%\rowcolor{lightgray}\multirow{4}{*} & \multicolumn{4}{l}{$\alpha =0.5$ and $a=0.2$} \\ \hline

\rowcolor{lightgray}{$\ell$} & $\omega _{\rm{WKB}}$ & $\omega _{\rm{S}}$ & $\Delta _{\rm{R}} \%$ & $\Delta _{\rm{I}} \%$ \\ \hline\hline
$5$ & 1.24912 - i 0.118982  & 1.1352 - i 0.118807  & 10.035 & 0.14730 \\ 
$7$ & 1.70307 - i 0.118901  & 1.5892 - i 0.118807   & 7.1652 &  7.9120 $\times 10^{-2}$ \\ 
$10$ & 2.38408 - i 0.118855 & 2.2703 - i 0.118807  & 5.0103 & 4.0402 $\times 10^{-2}$ \\ 
$15$ & 3.51917 - i 0.118829 & 3.4050 - i 0.118807    & 3.3530 & 1.8517 $\times 10^{-2}$ \\ 
$20$ & 4.65430 - i 0.118820   & 4.5407 - i 0.118807   & 2.5018 & 1.0942 $\times 10^{-2}$ \\
$30$ & 6.92459 - i 0.118813 & 6.8110 - i 0.118807   & 1.6677 & 5.0502 $\times 10^{-3}$\\
$40$ & 9.19490 - i 0.118810   & 9.0813 - i 0.118807  & 1.2509 & 2.5251 $\times 10^{-3}$\\
$50$ & 11.4652 - i 0.118809 & 11.352 - i 0.118807   & 0.99718 & 1.6834 $\times 10^{-3}$\\ \hline\hline
\end{tabular}
}
\caption{QNMs calculated by 6th-order WKB approximation method and by the shadow radius, where $a=0.2$ and $\alpha =0.5$.}
\label{tab:q2}
\end{table}

%%%%%%%%%%%%%%%%%%%%%%%%%%%%%%%%%%%%%%%%%%%%%%%%%%%%%%%%%%%%%%%%%%%%%%%
\begin{table}[H]
\centering{\normalsize \centering%
\begin{tabular}{l|l|l|ll}
\hline\hline
%\rowcolor{lightgray}\multirow{4}{*} & \multicolumn{4}{l}{$\alpha =0.1$ and $a=0.2$} \\ 
\hline

\rowcolor{lightgray}{$\ell$} & $\omega _{\rm{WKB}}$ & $\omega _{\rm{S}}$ & $\Delta _{\rm{R}} \%$ & $\Delta _{\rm{I}} \%$ \\ \hline\hline

$5$ & 0.636700 - i 0.0506996  & 0.578822 - i 0.0506555  & 9.9993 & 8.7059 $\times 10^{-2}$ \\ 
$7$ & 0.868229 - i 0.0506792 & 0.810350 - i 0.0506555 & 7.1425 &  4.6787 $\times 10^{-2}$\\ 
$10$ & 1.21552 - i 0.0506676&1.15764 - i 0.0506555 & 4.9998 &  2.3887 $\times 10^{-2}$ \\ 
$15$ & 1.79435 - i 0.0506610 & 1.73646 - i 0.0506555 & 3.3338 & 1.0858 $\times 10^{-2}$\\ 
$20$ & 2.37317 - i 0.0506587 & 2.31529 - i 0.0506555 & 2.4999 &  6.3172 $\times 10^{-3}$\\ 
$30$ & 3.53081 - i 0.0506569 & 3.47293 - i 0.0506555 & 1.6666 &  2.7638 $\times 10^{-3}$\\
$40$ & 4.68845 - i 0.0506563 & 4.63057 - i 0.0506555 & 1.2500 & 1.5793 $\times 10^{-3}$\\
$50$ & 5.84610 - i 0.0506560  & 5.78822 - i 0.0506555   & 0.99996 &  9.8706 $\times 10^{-4}$\\ 
\hline\hline
\end{tabular}
}
\caption{QNMs calculated by 6th-order WKB approximation method and by the shadow radius, where $a=0.2$ and $\alpha =-0.1$.}
\label{tab:q3}
\end{table}

%%%%%%%%%%%%%%%%%%%%%%%%%%%%%%%%%%%%%%%%%%%%%%%%%%%%%%%%%%%%%%%%%%
\begin{table}[tbh]
\centering{\normalsize \centering%
\begin{tabular}{l|l|l|ll}
\hline\hline
%\rowcolor{lightgray}\multirow{4}{*} & \multicolumn{4}{l}{$\alpha =0.1$ and $a=0.2$} \\ 
\hline

\rowcolor{lightgray}{$\ell$} &$\omega _{\rm{WKB}}$ & $\omega _{\rm{S}}$ & $\Delta _{\rm{R}} \%$ & $\Delta _{\rm{I}} \%$ \\ \hline\hline

$5$ & 0.468360 - i 0.0348526 & 0.425829 - i 0.0348283  & 9.9878 &  6.9771 $\times 10^{-2}$ \\ 
$7$ & 0.638706 - i 0.0348414 & 0.596161 - i 0.0348283 & 7.1365 & 3.7613 $\times 10^{-2}$\\ 
$10$ & 0.894216 - i 0.0348349 & 0.851658 - i 0.0348283 & 4.9971 &  1.8950 $\times 10^{-2}$\\ 
$15$ & 1.32006 - i 0.0348313 & 1.27749 - i 0.0348283  & 3.3323 & 8.6137 $\times 10^{-3}$\\ 
$20$ & 1.74589 - i 0.0348300 & 1.70332 - i 0.0348283    & 2.4992 &  4.8811 $\times 10^{-3}$\\ 
$30$ & 2.59756 - i 0.0348290 & 2.55497 - i 0.0348283 & 1.6669 & 2.0099 $\times 10^{-3}$\\
$40$ & 3.44922 - i 0.0348287 & 3.40663 - i 0.0348283  & 1.2502 & 1.1485 $\times 10^{-3}$\\
$50$ & 4.30088 - i 0.0348285 & 4.25829 - i 0.0348283  & 1.0002 & 5.7425 $\times 10^{-4}$\\ 
\hline\hline
\end{tabular}
}
\caption{QNMs calculated by 6th-order WKB approximation method and by the shadow radius, where $a=0.2$ and $\alpha =-0.5$.}
\label{tab:q4}
\end{table}

%\newpage
\noindent We observe that as {$\ell$} increases, the differences in both the real and imaginary components of the frequency gradually diminish. This trend suggests that the eikonal limit method is highly accurate for large angular momentum values, producing results that closely align with the WKB approximation. This finding holds significant practical value, as the eikonal limit method is generally more straightforward for analytical computations. Furthermore, high-precision results provide reliable data for investigating the stability and oscillatory behavior of black holes and other compact objects.

\section{Conclusion} \label{sec6}

Recent studies indicate that the properties of black holes can be significantly influenced by their surroundings, including factors such as thermal stability, $P-V$ criticality, quasinormal modes, shadow characteristics, and photon orbit radii. Inspired by this observation, we have investigated the complex relationship between the SBH and its environment, characterized by PFDM and a CoS. Our analysis shows that the presence of the CoS and PFDM modifies the metric function, thereby impacting the black hole's horizon and mass. Furthermore, our examination of the Hawking temperature reveals distinct behaviors based on the parameters of PFDM: for positive parameters, the temperature consistently decreases, while for negative parameters, it initially rises to a maximum before cooling down, especially when the CoS parameters are relatively small. Additionally, our analysis of specific heat illustrates how the perfect fluid dark matter parameters affect the black hole's stability; with a positive parameter, the heat capacity remains consistently negative, indicating instability, whereas a negative parameter allows the heat capacity to reach zero at a certain point, marking the minimum size at which the black hole ceases to radiate. Notably, the Gibbs free energy provides insights into the stability of the black hole. Although the surroundings do not directly alter the black hole's entropy, which remains linearly proportional to the area, they influence the horizon, which, in turn, affects the entropy. 

The geodesic structure and QNMs analysis reveal how the CoS and PFDM significantly impact the black hole’s dynamics and observable characteristics. To do this, we first derived the effective potential, and then, we investigated the stability of particle orbits, particularly altering the photon sphere and innermost stable circular orbit. We demonstrated that these changes have direct implications for the black hole's shadow radius. In other words, the study of null and timelike geodesics shows that the parameters of the CoS and PFDM strongly influence the trajectories of light and matter around the black hole. Furthermore, the quasinormal mode analysis, using both the WKB approximation and shadow radius methods, indicated that the CoS and PFDM not only affect the oscillation frequencies but also the decay rates of the black hole’s perturbative modes. The real part of the QNM frequency is found to be linked to the angular velocity of the photon sphere, while the imaginary component is tied to the Lyapunov exponent, which governs the rate of mode decay. These results suggest that the presence of CoS and PFDM can provide distinctive signatures in the black hole’s radiative and oscillatory behavior, potentially observable through gravitational wave detections. Such findings enhance our understanding of how exotic matter fields influence the stability, dynamics, and radiative properties of black holes, offering new pathways for testing theoretical models of dark matter and string clouds in astrophysical contexts.

{ Finally, it is important to emphasize that black holes should not be regarded as isolated entities. In nature, they constantly interact with their surroundings and are rarely unaffected by neighboring fields. Consequently, they are typically in a perturbed state due to the influence of other fields in their vicinity. The presence of PFDM and CoS around black holes offers valuable opportunities to test and expand gravity theories. PFDM modifies black hole thermodynamics and particle orbits, influencing key properties such as temperature, phase transitions, and gravitational lensing. These effects establish a bridge between the cosmological role of dark matter and its impact on black hole physics. On the other hand, CoS affects the geometry of spacetime and the physical phenomena nearby. The mathematical structure of string theory has consistently demonstrated that it encompasses all the essential components for a quantum theory of gravity, including valuable insights into black hole entropy. However, string theory is not merely a theory of quantum gravity; it also incorporates particles and interactions, providing the tools necessary to address contemporary questions in cosmology. PFDM specifically addresses the coupling between dark matter and gravity, while the effects of CoS provide concrete tests of string theory-inspired modifications and potential signatures of quantum gravity. These predictions are particularly significant, as they can be tested using current and future observational facilities, such as LIGO/Virgo, the Event Horizon Telescope, and X-ray observatories. }

\section*{Acknowledgments}
%We are thankful to the editor and anonymous referees for their constructive suggestions and comments. This work is supported by the Ministry of Higher Education and Scientific Research, Algeria under the code: B00L02UN040120230003. 
{ We would like to express our sincere gratitude to the anonymous referee for their constructive suggestions and comments.} B. C. L. is grateful to Excellence project PřF UHK 2211/2023-2024 for the financial support.

\end{document}